\documentclass[sn-mathphys]{sn-jnl}

\theoremstyle{thmstyleone}%
\theoremstyle{thmstyletwo}%

\theoremstyle{thmstylethree}%

\begin{document}

\title[Inventory of high-quality flat-band van der Waals materials]{Inventory of high-quality flat-band van der Waals materials}


\author[1,2,5]{\fnm{Jingyi} \sur{Duan}}\email{duanjy0518@gmail.com}

\author*[3,1,2]{\fnm{Da-Shuai} \sur{Ma}}\email{mads@cqu.edu.cn}

\author*[1,2]{\fnm{Run-Wu} \sur{Zhang}}\email{zhangrunwu@163.com}

\author[4]{\fnm{Zeying} \sur{Zhang}}\email{zzy@mail.buct.edu.cn}

\author[1,2]{\fnm{Chaoxi} \sur{Cui}}\email{cuichaoxi@bit.edu.cn}

\author[1,2]{\fnm{Wei} \sur{Jiang}}\email{jiangw@bit.edu.cn}

\author[1,2]{\fnm{Zhi-Ming} \sur{Yu}}\email{zhiming\_yu@bit.edu.cn}

\author*[1,2]{\fnm{Yugui} \sur{Yao}}\email{ygyao@bit.edu.cn}

\affil[1]{\orgdiv{Centre for Quantum Physics, Key Laboratory of Advanced Optoelectronic Quantum Architecture and Measurement (MOE), School of Physics}, \orgname{Beijing Institute of Technology}, \orgaddress{ \city{Beijing}, \postcode{100081},  \country{China}}}

\affil[2]{\orgdiv{Beijing Key Lab of Nanophotonics and Ultrafine Optoelectronic Systems, School of Physics},  \orgname{Beijing Institute of Technology}, \orgaddress{ \city{Beijing}, \postcode{100081},  \country{China}}}

\affil[3]{\orgdiv{Institute for Structure and Function \& Department of Physics}, \orgname{Chongqing University}, \orgaddress{ \city{Chongqing}, \postcode{400044},  \country{China}}}

\affil[4]{\orgdiv{College of Mathematics and Physics}, \orgname{Beijing University of Chemical Technology}, \orgaddress{ \city{Beijing}, \postcode{100029},  \country{China}}}

\affil[5]{\orgdiv{College of Physics and Optoelectronic Engineering}, \orgname{Shenzhen University}, \orgaddress{ \city{Shenzhen}, \postcode{518060}, \country{China}}}

\abstract{
More is left to do in the field of flat bands besides proposing theoretical models. One unexplored area is the flat bands featured in  the van der Waals (vdW) materials. 
Exploring more flat-band material candidates and moving the promising materials toward applications have been well recognized as the cornerstones for the next-generation high-efficiency devices. 
Here, we utilize a powerful high-throughput tool to screen desired vdW materials based on the Inorganic Crystal Structure Database. 
Through layers of filtration, we obtained 861 potential monolayers from 4997 vdW materials. 
Significantly, it is the first example to introduce flat-band electronic properties in the vdW materials and propose three families of representative flat-band materials by mapping two-dimensional (2D) flat-band lattice models. 
Unlike existing screening schemes, a simple, universal rule, \textit{i.e.}, 2D flat-band score criterion, is first proposed to efficiently identify 229 high-quality flat-band candidates, and guidance is provided to diagnose the quality of 2D flat bands. 
All these efforts to screen experimental available flat-band candidates will certainly motivate continuing exploration towards the realization of this class of special materials and their applications in material science.
}




\maketitle

\section{Introduction}\label{sec1}

Flat bands, characterized by the high density of states and quenched electronic kinetic energy, are considered as desired paradigm to explore fascinating physical phenomena such as superconductivity\cite{white_electronic_1981,sutherland_simple_1986,lieb_two_1989,Kopnin2011,tang_strain-induced_2014,iglovikov_superconducting_2014,volovik_graphite_2018,yin_negative_2019,liu_spectroscopy_2021}, charge-density-wave state\cite{rice_new_1975,carpinelli_direct_1996,calandra_phonon-assisted_2018,wang_charge_2020}, and nontrivial band topology\cite{miyahara_bcs_2007,dora_lattice_2011,wang_nearly_2011,bergholtz_topological_2013,mukherjee_observation_2015,morales-inostroza_simple_2016,julku_geometric_2016,mondaini_pairing_2018,maimaiti_compact_2017,leykam_artificial_2018,gardenier_p_2020,chiu_fragile_2020,ma_spin-orbit-induced_2020,cadez_metal-insulator_2021}. 
The recently discovered three-dimensional (3D) compounds\cite{provenzano_bringing_2020,ghimire_competing_2020,lou_charge-density-wave-induced_2022,neupert_charge_2022} with  kagome lattice composed of transition-metal ions surprisingly show the coexistence of all of the intriguing phenomena, which has attracted broad interest in condensed-matter physics and material science. 
As these materials share kagome geometries\cite{mielke_ferromagnetism_1991,mielke_ferromagnetic_1991,chiu_fragile_2020,ma_spin-orbit-induced_2020}, it is naturally expected to host those well-known features of such a unique lattice model, such as flat bands and Dirac fermions. However, due to the lack of an ideal candidate that has suitable electron filling and clear kagome bands, and the complicated interactions among electrons from various sublattices, these remarkable characteristics of flat-band effects or Dirac fermions are yet been observed in the 3D kagome compounds that have a perfect geometric kagome lattice. Therefore, it remains essential to search for high-quality materials with various flat-band lattice models to pave the way for exploring exotic correlation effect due to flat band through materials characterizations.

\hspace*{\fill}

\begin{figure}
	\includegraphics[width=\linewidth]{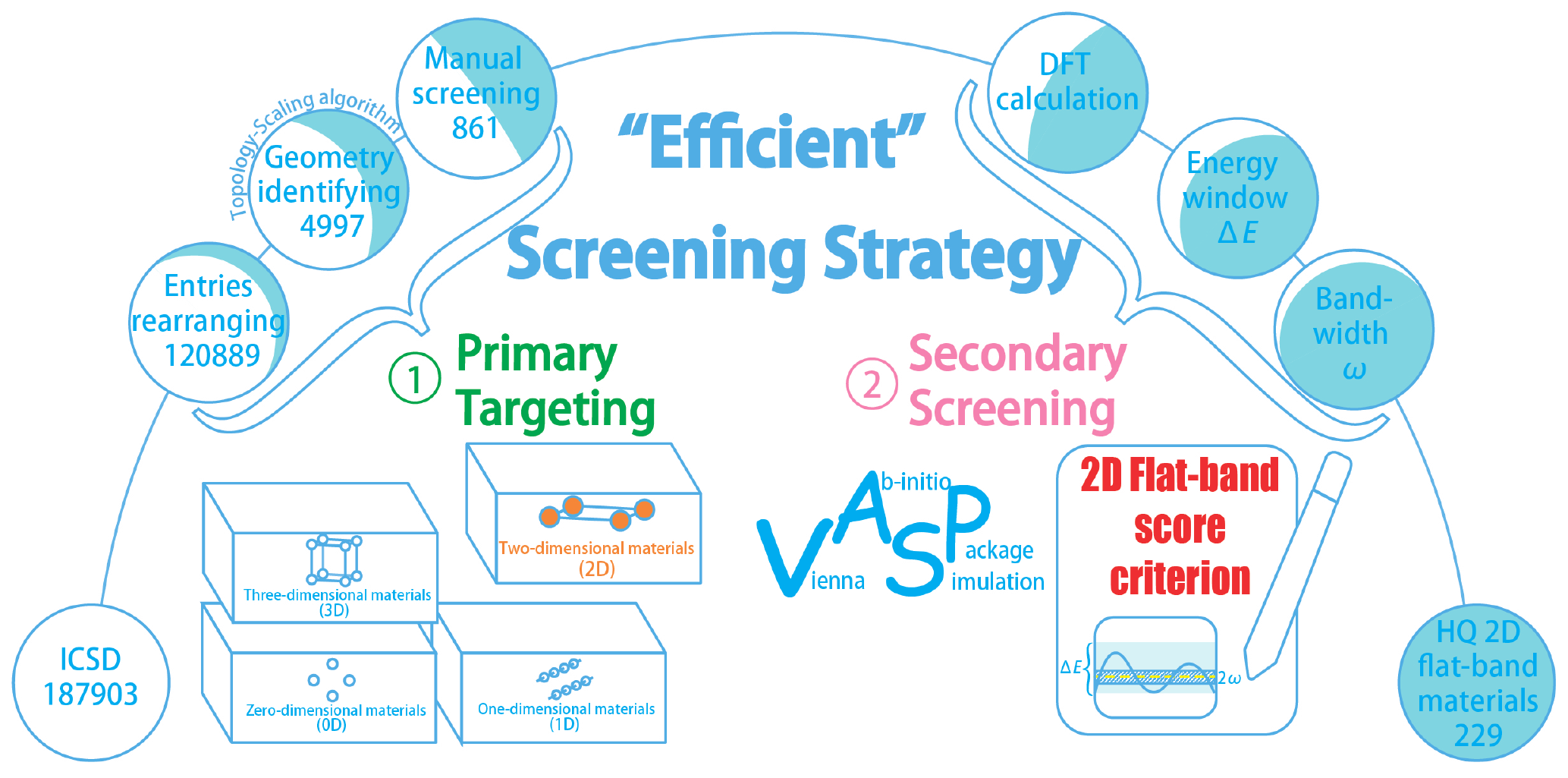}
	\caption{The schematic flowchart of the inventory of van der Waals flat-band materials. This workflow can identify the features of flat bands and illustrate the details of constructing a high-quality (HQ) \textit{Flat-band Materials Package} by two tiers:  \textit{primary targeting} and \textit{secondary screening}.}
	\label{fig:boat1}
\end{figure}

Motivated by the unique advantages of two-dimensional (2D) materials\cite{chen_two-dimensional_2018,jin_emerging_2018,han_van_2018,zeng_exploring_2018,zhang_introduction_2018,dong_interface-assisted_2018,cai_chemical_2018,li_epitaxial_2018}, more recent studies of flat-band effects has been focusing on 2D systems, especially van der Waals (vdW) materials, rather than its 3D counterparts\cite{jacqmin_direct_2014,wang_flatbands_2016,zhang_flat_2019,crasto_de_lima_topological_2019,maimaiti_flat-band_2021}. Moreover, the desire to miniaturize future quantum devices has stimulated great research interest in vdW materials. Known vdW materials like graphene\cite{geim_graphene_2009}  are characterized by weak interlayer interactions through vdW forces, implying that such materials could be easily exfoliated from parent 3D crystals. It presents excellent opportunities to further explore unexplored areas of physical properties\cite{guo_graphene_2011,lazar_adsorption_2013,kong_molecular_2014,si_strain_2016} that are constrained in 3D crystals and only possible for 2D layered systems. As a prominent example, the recently reported twisted bilayer graphene\cite{cao2018unconventional,PhysRevLett2018.121.087001,PhysRevB.98.085435,PhysRevB.98.085436,PhysRevX2018.8.031089,PhysRevX2018.8.041041,PhysRevLett2018.121.257001,laksono2018singlet,PhysRevLett.121.217001,PhysRevB.98.195101,PhysRevB.98.220504,PhysRevB.98.241407,guinea2018electrostatic,PhysRevB.99.121407,PhysRevLett.122.257002,PhysRevLett.122.026801,PhysRevLett.122.246402,yankowitz2019tuning,lu2019superconductors,sharpe2019emergent,huang2019antiferromagnetically,wu2018emergent} shows  topologically nontrivial flat bands  at the magic twisting angle, yielding the hallmarks of electron correlations, magnetism as well as unconventional superconductivity, which has spurred rapid development in twistronics. Compared to 3D crystals, the 2D vdW candidates, with lower dimensional, could not only easily map out obvious flat-band lattice models but also provide more straightforward visual evidence to capture their prime features.  Since vdW materials are potentially applicable for the study of flat-band effects, it is urgent to find desired realistic vdW crystals with high-quality flat band through a simple and effective approach, which could help researchers further comprehend the inherent nature of strong electron interactions.

\hspace*{\fill}

Here, we skillfully performed high-throughput calculations to screen vdW materials with high-quality flat bands from the Inorganic Crystal Structure Database (ICSD)\cite{belsky_new_2002}. Starting from the 861 materials that can be easily exfoliated through the layers of filtration, we identify a subset of 229 high-quality flat-band cases according to a simple-to-use flat-band score criterion. In particular, the selected candidates display abundant structural prototypes, simple and feasible material components, covering manifold features such as magnetism, topology and superconductivity induced by the inherently strong electron interactions. Furthermore, several intriguing flat-band lattice models are presented based on selected flat-band vdW materials, including breathing kagome lattice, honeycomb lattice, and twisted-kagome lattice. Overall, the 2D high-quality flat-band candidates chosen by our research can shed new light to the intrinsic 2D flat-band physics.

\section{Results}\label{sec2}

The main results of the inventory are summarized as follows. We first propose an efficient two-tier screening strategy: \textit{primary targeting} and \textit{secondary screening}. The former screening process mainly focuses on the selection of potential 2D candidates, \textit{i.e.}, vdW materials, from the material database to pave the way for the deeper processing of the latter. Secondary screening aims to filter high-quality flat-band materials by first-principles calculations and our newly proposed 2D flat-band score criterion. After that, we classify the selected materials based on those intriguing properties that of general interest, including geometrical characteristics, symmetry information, conductivity, and magnetism, which could provide valuable guidance for further research. Last but not least, we showcase the three best representative flat-band materials by mapping their electronic band structure and further develop flat-band effective lattice models based on structural symmetry to unravel the main physics of the flat band.

\begin{figure}
	\includegraphics[width=\linewidth]{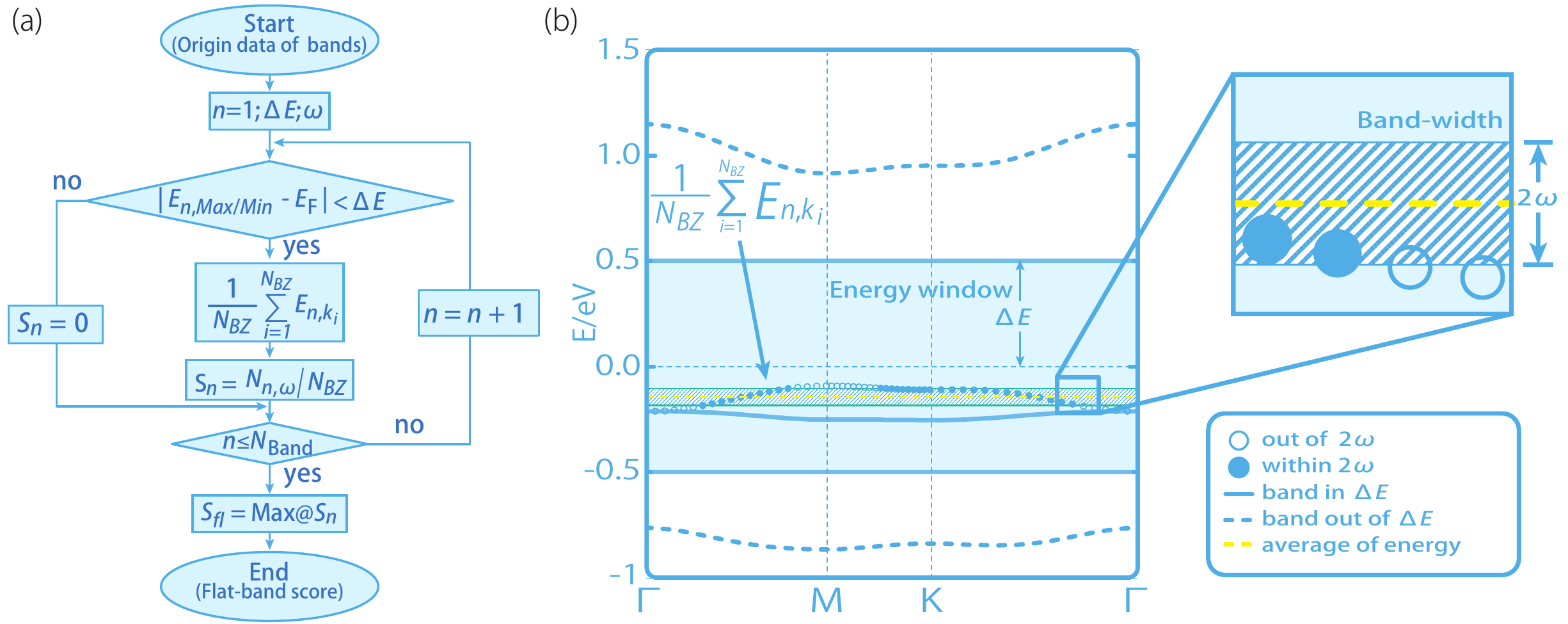}
	\caption{ (a) Workflow to construct the 2D flat-band score criterion; the core part is to determine the energy window $\Delta E$ and the band-width $\omega$. (b) Graphical representation of the algorithm fornumerically defining the flatness of band structure. Here we show the evaluation process when  $\Delta E$ = 0.5 eV and $\omega$ = 0.025 eV. The light blue area represents the area covered by the energy window $\Delta E$, and only the band that is wholly locating in the energy windows  $\Delta E$ is scored separately. The yellow dashed line represents the average value of the energy of the band plotted in blue soild dots and blue hollow dots, and the blue and white diagonal striped area represents the effective area defined by the expansion of the yellow dashed line along with the ±$\omega$.}
	\label{fig:boat2}
\end{figure}

\subsection{Efficient screening strategy}\label{subsec1}

There are two tiers in the high-quality vdW flat-band materials screening process: \textit{primary targeting} (first-tier raw material targeting) and \textit{secondary screening} (second-tier fine feature screening), as sketched in Fig.~\ref{fig:boat1}. Primary targeting can be further divided into three parts, namely entries rearranging, geometry identifying, and manual screening, as shown in Fig.~\ref{fig:boat1} (left panel). We first excluded the seldom studied entries of the ICSD: radioactive elements, noble gas elements, elements with atomic number greater than 93, and the entries with a total number of atoms that exceed 200. The number of targets dropped sharply from 187093 to 120889. After that, we adopted geometry identifying method to obtain vdW materials from those normal candidates. 
To explore the relationship between material dimensionality and the size of covalently bonded clusters, we adopt a topology-scaling algorithm\cite{ashton_topology-scaling_2017,cheon_data_2017,mounet_two-dimensional_2018,zhu_systematic_2018} which could help us uncover the 2D structures. The results yield 4997 possible layered materials to be promising monolayers through feasible exfoliation. We finally establish a portfolio containing 861 candidates by further eliminating defective entries, and each case has a corresponding ICSD number. The details of our screening criteria are shown in the section I of \textit{Supplementary Materials} (SM).

\hspace*{\fill}

The secondary fine feature screening, as illustrated in Fig.~\ref{fig:boat1} (right panel), is performed following first-principles calculations and 2D flat-band score criterion. In order to better extract the flat-band effective lattice model, we only focus on the cases without spin-orbit coupling. During the screening, we first collate the structural parameters and the electronic properties (band structure) of 861 candidates featuring vdW characteristics, which provides necessary guidelines for the high-throughput screening of flat-band materials. One bottleneck of current flat-band fields is to efficiently evaluate the quality of flat-band materials that can be help to advance the feasible experiments. Therefore, we further introduce a simple, universal rule for evaluating band flatness, \textit{i.e.}, 2D flat-band score criterion, to automatically evaluate the quality of flat-band materials among 861 candidates.

\hspace*{\fill}

To characterized the band flatness, we implement a quantitative approach to automatically search for flat-band candidates. The flat-band score for the studied material ($S_{fl}$) is a function of energy window ($\Delta E$), band-width ($\omega$), and the total number of points ($N_{\mathrm{BZ}}$) along the calculated high-symmetry lines in the first Brillouin Zone (BZ). The flow chart of getting the flat-band score $E_{\text{fl}}$ is shown in Fig.~\ref{fig:boat2} (a). More specifically, it is for materials with $N_{\text{fl}}$ bands that are totally located within the energy window $ \lvert E_{n,\text{Min/Max}} - E_{F}\rvert< \Delta E $. The flat-band score of each band ($S_{n}$, $n$ is the band index) indicates, essentially, the ratio of the number of the $k$-points with energy $E_{n,\boldsymbol{k}}$ that satisfy $\lvert E_{n,\boldsymbol{k}}-\frac{1}{N_{\mathrm{BZ}}}\sum_{i=1}^{N_{\mathrm{BZ}}}E_{n,\boldsymbol{k}_{i}} \rvert < \omega$  to the total $k$-points number $N_{\mathrm{BZ}}$, which ranges from 0 to 1. 
Lastly, the flat-band score of the material, $S_{fl}$, is determined as the maximum of $S_{n}$ in the energy window $\Delta\textit{E}$, as sketched in Fig.~\ref{fig:boat2}, which has the same range from 0 to 1 as individual band score. If a score is greater than 0, there is at least one flat band segment in the studied band structure section. For the extreme scenario, a score of 1 suggests that the candidate possesses a perfectly flat band under the prescribed conditions.

\begin{figure}
	\includegraphics[width=\linewidth]{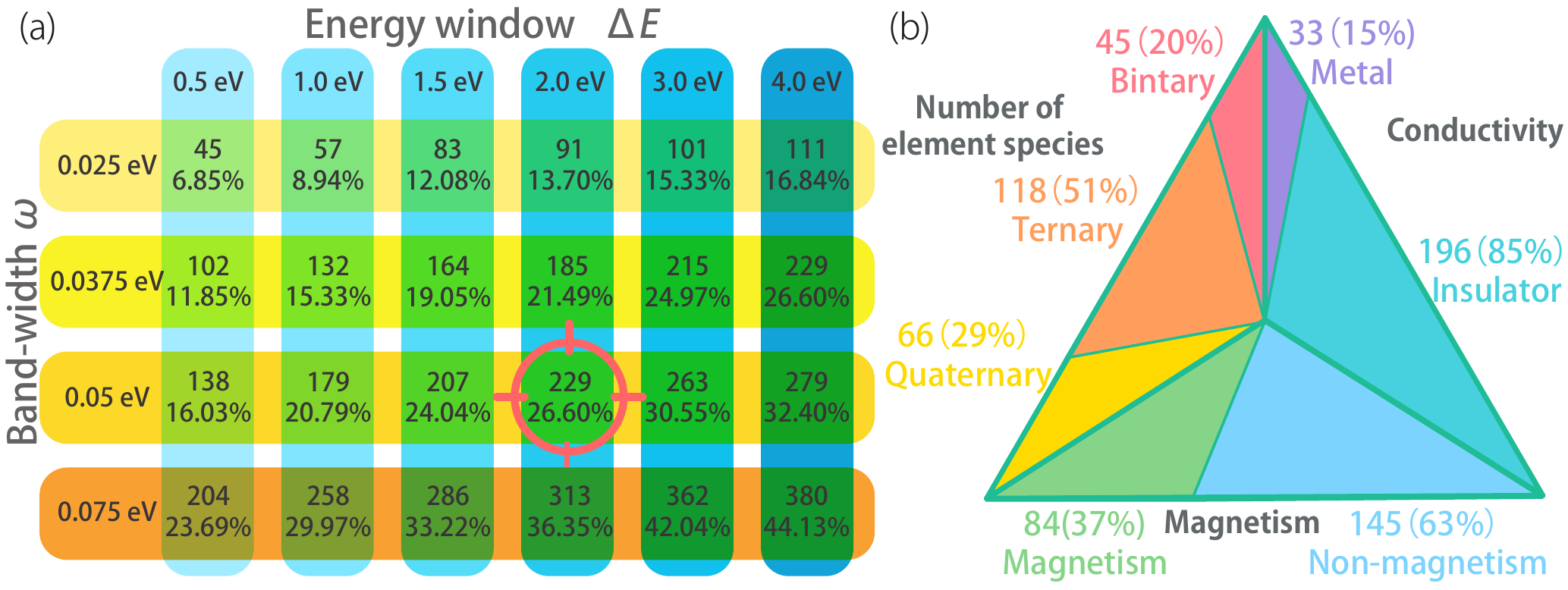}
	\caption{(a) Distribution of high-quality flat-band materials discrete values of $ \omega $ and $\Delta E$.  (b) The number of element species, conductivity and magnetism of the high-quality flat-band materials obtained by  $\omega$ = 0.05~eV and  $\Delta E$ = 2~eV are adopted.}
\label{fig:boat3}
\end{figure}

\subsection{Portfolio of high-quality screened materials}\label{subsec2}

With the efficient screening strategy, we have identified an extensive portfolio of high-quality monolayer structures by setting the reasonable range of $\Delta E$ and $\omega$, as summarized in Fig. $\ref{fig:boat3}$ (a). In this study, a combination of $\Delta E$, $\omega$ (2 eV, 0.05 eV) is chosen as the flat-band score criterion, which searches for bands within a small energy window around the Fermi level with a reasonable small band dispersion. Without loss of generality, the results of other screening criteria are tabulated in Fig. \ref{fig:boat3} (b). Under this screening criterion, there are 229 high-quality flat-band materials that have a flat-band score greater than 0.9. To facilitate potential future researches, the 229 high-quality flat-band monolayers are further classified based on the number of elements, conductivity, and magnetism, as summarized in Fig. \ref{fig:boat3} (b). To improve the experimental feasibility, many structures are listed as candidates, including 45 binary, 118 ternary, and 66 quaternary compounds. By exploring the band structures of 229 screened materials, there are 145 (84) nonmagnetic (magnetic) materials and 196 (33) insulators (metals). The band structures of all high-quality flat-band materials are displayed in SM. Many flat bands can be easily explained theoretically via mapping flat-band lattices. This portfolio of high-quality flat-band materials offers a unique, easy-to-implement, and robust platform for exploring various intriguing properties of flat-band system.

\subsection{Mapping flat-band lattices}\label{subsec3}
Leveraging the 2D candidates with atomic scale, one can provide more visual evidence to capture flat-band features by mapping clearer flat-band lattices compared to the 3D system. To illustrate this, we demonstrate three representative examples from our selected high-quality flat-band materials, such as breathing-kagome lattice, honeycomb lattice, and twisted-kagome lattice to further display the physical origin of the flat bands.

\hspace*{\fill}

The newly developed $S$-matrix theory\cite{calugaru_general_2022} draws out the connection between the lattices' simple geometric features, the emergence of flat bands, and the band representation of the potential flat bands. In the $S$-matrix theory, a lattice is formed by two sublattices labeled by L and L$'$, respectively. There are $N_{L}$ and $N_{L'}$ ($N_{L'}$ $<$ $N_{L}$) orbitals per unit cell. Suppose the band representations BR and BR$'$ (formed by orbitals in L and L$'$ sublattices) satisfy that BR-BR$'$ is inconsistent with any band representation, the flat bands must be topological. According to the $S$-matrix theory, breathing-kagome lattice, honeycomb lattice, and twisted-kagome lattice have precise flat bands, which are topologically nontrivial.

\hspace*{\fill}

From the perspective of real space, there is also one figurative understanding of the origin of the flat band in lattices models. It has been proved that we can always find compact localized states in real space to describe the flat band in breathing-kagome,  honeycomb, and twisted-kagome lattices (considering only the nearest neighbor couplings). In these lattices, the number of the real space eigenstates of each flat band is $N_{\text{orbital}}+1$, which indicates that the flat band touches with the dispersive bands at one point in BZ\cite{bergman_band_2008}. Based on the tight-binding (TB) Hamiltonian, more detailed explanations about the three kinds of flat-band systems are illustrated in the section III of SM.

\begin{figure}[h]
	\includegraphics[width=1\linewidth]{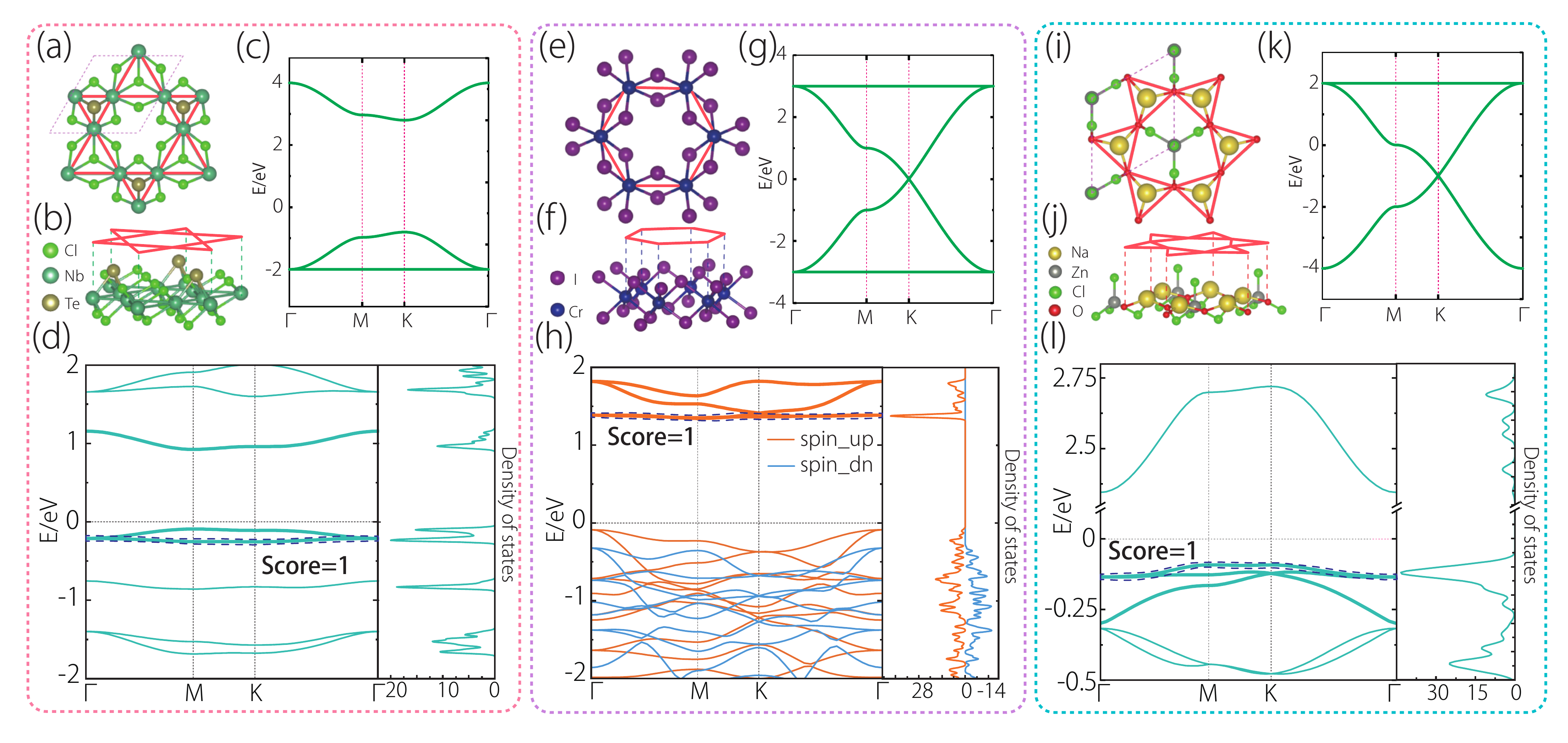}
	\caption{The geometries of crystals and band structures in the following tangible flat-band materials. Three representative flat-band lattices involve (a, b) breathing-kagome lattice, (e, f) honeycomb lattice, and (i, j) twisted-kagome lattice. Moreover, the corresponding TB model band structure are plotted in (c, g, k), respectively. For increased visibility, prominent flat-band lattices are marked by red lines. The concrete materials include: (a, b)  Nb$_{3}$TeCl$_{7}$ (icsd\_079213) sharing SG 156 ($P$3m1) are identified as a breathing-kagome lattice with the Nb atoms. (d) band structure and density of states of Nb$_{3}$TeCl$_{7}$ are calculated. Similarly, (e, f)  CrI$_{3}$ (icsd\_251655) featuring SG 162 ($P\bar{3}1m$) are identified as a honeycomb lattice with the Cr atoms, and band structure and density of states are shown in (h). The orange line represents the spin-up channel, and the blue line represents the spin-down channel; (i, j) Na$_{2}$ZnO$_{3}$Cl$_{4}$ (icsd\_024269) characterizing SG 157 ($P$31m) are identified as a twisted-kagome lattice with the O atoms, and band structure and density of states are plotted in (l).}
\label{fig:boat4}
\end{figure}

\subsubsection{Breathing-kagome lattice}\label{subsubsec2}

Among the identified 229 high-quality flat-band materials, it is worth noting that one class is found to exhibit the characteristics of a breathing-kagome lattice. For such type of flat-band lattice, we take Nb$_{3}$TeCl$_{7}$ as an example to explain the origin of the flat band. Since Te occupies one of the Cl hollow positions in Nb$_{3}$Cl$_{8}$, the Nb-atoms network forms a breathing-kagome structure composed of alternating smaller downward pointing equilateral triangles and bigger upward pointing equilateral triangles, are sketched in Figs.~\ref{fig:boat4} (a) and (b). In order to comprehend the key physics underlying the breathing-kagome lattice, we develop a simple three-band TB model for describing the flat-band characteristic [Fig.~\ref{fig:boat4} (c)]. To better test our theoretical analysis, the band structure of Nb$_3$TeCl$_7$ is obtained from the DFT calculations [Fig.~\ref{fig:boat4} (d)], which agrees well with the one of TB model. The related three bands, as highlighted in bold in Fig.~\ref{fig:boat4} (d), mainly arise from the $d_{z^2}$ orbitals of Nb, which further confirms that the flat band is indeed originated from the breathing-kagome lattice. The detailed Hamiltonians of Nb$_{3}$TeCl$_{7}$ is derived in the section III of SM.

\hspace*{\fill}

Besides Nb$_{3}$TeCl$_{7}$, the other M$_{3}$QX$_{7}$ family materials (M = Nb, Ta; Q = S, Se, Te; X = Cl, Br, I) also form the breathing-kagome lattice, which hosts a similar flat band. Furthermore, the breathing-kagome lattice contains not only non-magnetic vdW materials but also magnetic materials, such as Nb$_{3}$Cl$_{8}$, Nb$_{3}$Br$_{8}$ and Nb$_{3}$I$_{8}$, which could further help to reveal exotic physics of the flat band that is entangled with magnetism. More flat-band related information such as crystal structures and band structures is detailed in the section III of SM.

\hspace*{\fill}

\subsubsection{Honeycomb lattice}\label{subsubsec2}

One of the simplest and most charming flat-band lattice is the honeycomb lattice, as diagrammed in Fig.~\ref{fig:boat4} (e)-(h), which has triggered many profound physics studies. When placing t he \textit{p}$_{x}$ and \textit{p}$_{y}$ (or \textit{d}$_{yz}$ and \textit{d}$_{xz}$) atomic orbitals on the honeycomb lattice, it forms two flat bands in the band structure as shown in Fig.~\ref{fig:boat4} (g). Significantly, the flat bands in many high-quality flat-band materials are derived from the $p$/$d$-band honeycomb lattice, such as the known 2D CrX$_{3}$  family with X =Cl, Br, I. As shown in Fig.~\ref{fig:boat4}(e), CrI$_{3}$ shares a honeycomb lattice where two identical sublattices are formed by the Cr atoms. Each type of Cr atom is surrounded by an octahedron formed by six iodine atoms [see Figs.~\ref{fig:boat4} (e) and (f)]. With the orbital analysis, the set of bands are mainly composed by the \textit{d}$_{yz}$ and \textit{d}$_{xz}$ atomic orbitals of Cr atoms that are located on the honeycomb lattice. The flat band near 1.5 eV [(Fig.~\ref{fig:boat4} (h))] is in agreement with the flat band obtained by the TB model as shown in Fig.~\ref{fig:boat4} (g). Moreover, the similar flat band can be found in  CrBr$_{3}$ and CrCl$_{3}$. It is worth noting that the doping of alkali metal atoms in the vacancies of CrI$_{3}$ and CrBr$_{3}$ can adjust the relative position of the flat bands to the Fermi level\cite{xu_theoretical_2020,zhang_high-temperature_2020}. Thus, it is still valuable to study the properties of the flat band in CrX$_{3}$ although the flat band is intrinsically away from the Fermi level. The crystal structures and band structures of the other CrX$_{3}$ materials are given in the section IV of SM.

\subsubsection{Twisted-kagome lattice}\label{subsubsec2}

As a deformed kagome lattice, twisted-kagome lattice is characterized by rotating adjacent site-sharing triangles in opposite directions, as is shown in Figs.~\ref{fig:boat4} (i) and (j). Considering only the nearest neighbor couplings, there are a set of flat bands in the band structure of twisted-kagome lattice while putting $s$ orbital or hybrid $p$ orbitals on the occupied sites, as plotted in Fig.~\ref{fig:boat4}(k). We find that such lattice widely exists in our identified high-quality flat-band materials. Here, we describe one possible existence of flat band in a hitherto unrecognized material Na$_{2}$ZnO$_{3}$Cl$_{4}$. The crystal structure of Na$_{2}$ZnO$_{3}$Cl$_{4}$ is illustrated in Fig.~\ref{fig:boat4}(i) and (j), where the O atoms form a twisted-kagome lattice. The \textit{p}$_{x}$ and \textit{p}$_{y}$ orbitals of O atoms contribute to the flat band near the Fermi level [Fig.~\ref{fig:boat4} (l)]. The TB model and the schematic diagram of the structure and electronic properties of this material are given in the section III of SM.

\section{Discussion and Conclusion}\label{sec3}

\textit{Screened materials are real and feasible.}
Distinct from the existing high throughput research on flat bands\cite{regnault_catalogue_2022,PhysRevMaterials.5.084203}, we propose for the first time that the vdW materials are most likely to achieve flat-band characteristics. Such flat-band materials are easy to synthesize, peel, and transfer, which paves the way for future experimental studies. It is expected that the vdW materials with flat bands will provide a solid foundation for the exploration of exotic quantum phenomena (\textit{e.g.}, such as electron correlations, magnetism, and superconductivity) and the intriguing applications of topological quantum computation in future quantum devices. Among the 229 high-quality flat-band vdW materials, many are experimentally synthesized materials with their flat band properties overlooked that awaits further experimental investigation. For example, the bulk compound Nb$_{3}$TeCl$_{7}$ has been synthesized and structurally characterized by single-crystal X-ray diffraction in 1995\cite{miller_solid_1995}. In such lattice, Nb atoms that contribute to the flat band are distorted from high-symmetry locations to produce trimers with shorter interatomic distances, which forms the basis of ferroelectric polarization. Lattice breathing interchanges trimer patterns that further switches the direction of ferroelectric polarization, which simultaneously remodels the flat band. Another example is CrI$_{3}$, which has been reported as intrinsic ferromagnetism in atomically thin crystals. The family of CrX$_{3}$ (X=Cl, Br, I) materials exhibit plentiful intriguing properties\cite{huang_layer-dependent_2017,lin_critical_2018,shcherbakov_raman_2018,thiel_probing_2019} the giant tunneling magnetoresistance, gate tunable magneto-optical Kerr effect, and a large magnetocrystalline anisotropy. Screening results showcase that flat band vdW materials are abundant, implying the great potential for realizing flat band characteristics of different realistic applications. Future experiments could more easily access these flat bands with a high density of states, which renders more exciting physics phenomena, such as superconductivity, charge-density-wave state, and nontrivial band topology.

\hspace*{\fill}

\textit{Filtering strategies are unique and efficient.}
The recently discovered potential flat-band compounds featured by the kagome lattice composed of transition-metal ions have attracted much theoretical and experimental attention. Because of the kagome lattice of these materials, it was touted as a promising platform to detect the flat band. However, it is far more challenging to characterize the flat bands due to the lack of ideal candidates and/or vanishing flat-band feature due to many-body interaction though with a kagome lattice. It remains an open question whether one can find a convenient indicator to evaluate the flatness of the band. Once identified, such  indicator can help evaluate the quality of flat bands. In this work, a simple, universal rule, \textit{i.e.}, 2D flat-band score criterion, is proposed to identify various flat-band candidates efficiently, and guidance is provided to diagnose the quality of 2D flat bands. Significantly, our criterion can quantitatively determine the flatness of the screened vdW materials, which provide over 229 high-quality candidate materials for experimental verification.

\hspace*{\fill}

In this work, we have performed a systematic search for flat-band vdW materials and further explored the flat-band physical mechanism by mapping the types of lattices from the vdW candidates. By implementing effective screening strategies to 861 unique monolayers, we found that 229 candidates (score criterion: $\omega$ = 0.05 eV and $\Delta E$ = 2 eV) host high-quality flat bands. The appearance of flat bands in materials can be, in large but non-exhaustive part, theoretically understood using the $S$-matrix method, as we have exemplified in three representative candidates, \textit{i.e.}, Nb$_{3}$TeCl$_{7}$, CrI$_{3}$ and Na$_{2}$ZnO$_{3}$Cl$_{4}$. Our results work as a guide for future theoretical and experimental studies on 2D flat-band materials where exotic physics phenomena such as magnetism and superconductivity can be further explored. All the results obtained in this work and detailed in the SM can be accessed on the \textit{Flat-band Materials Package}.


\bibliography{ref}


\begin{thebibliography}{96}
\ifx \bisbn   \undefined \def \bisbn  #1{ISBN #1}\fi
\ifx \binits  \undefined \def \binits#1{#1}\fi
\ifx \bauthor  \undefined \def \bauthor#1{#1}\fi
\ifx \batitle  \undefined \def \batitle#1{#1}\fi
\ifx \bjtitle  \undefined \def \bjtitle#1{#1}\fi
\ifx \bvolume  \undefined \def \bvolume#1{\textbf{#1}}\fi
\ifx \byear  \undefined \def \byear#1{#1}\fi
\ifx \bissue  \undefined \def \bissue#1{#1}\fi
\ifx \bfpage  \undefined \def \bfpage#1{#1}\fi
\ifx \blpage  \undefined \def \blpage #1{#1}\fi
\ifx \burl  \undefined \def \burl#1{\textsf{#1}}\fi
\ifx \doiurl  \undefined \def \doiurl#1{\url{https://doi.org/#1}}\fi
\ifx \betal  \undefined \def \betal{\textit{et al.}}\fi
\ifx \binstitute  \undefined \def \binstitute#1{#1}\fi
\ifx \binstitutionaled  \undefined \def \binstitutionaled#1{#1}\fi
\ifx \bctitle  \undefined \def \bctitle#1{#1}\fi
\ifx \beditor  \undefined \def \beditor#1{#1}\fi
\ifx \bpublisher  \undefined \def \bpublisher#1{#1}\fi
\ifx \bbtitle  \undefined \def \bbtitle#1{#1}\fi
\ifx \bedition  \undefined \def \bedition#1{#1}\fi
\ifx \bseriesno  \undefined \def \bseriesno#1{#1}\fi
\ifx \blocation  \undefined \def \blocation#1{#1}\fi
\ifx \bsertitle  \undefined \def \bsertitle#1{#1}\fi
\ifx \bsnm \undefined \def \bsnm#1{#1}\fi
\ifx \bsuffix \undefined \def \bsuffix#1{#1}\fi
\ifx \bparticle \undefined \def \bparticle#1{#1}\fi
\ifx \barticle \undefined \def \barticle#1{#1}\fi
\bibcommenthead
\ifx \bconfdate \undefined \def \bconfdate #1{#1}\fi
\ifx \botherref \undefined \def \botherref #1{#1}\fi
\ifx \url \undefined \def \url#1{\textsf{#1}}\fi
\ifx \bchapter \undefined \def \bchapter#1{#1}\fi
\ifx \bbook \undefined \def \bbook#1{#1}\fi
\ifx \bcomment \undefined \def \bcomment#1{#1}\fi
\ifx \oauthor \undefined \def \oauthor#1{#1}\fi
\ifx \citeauthoryear \undefined \def \citeauthoryear#1{#1}\fi
\ifx \endbibitem  \undefined \def \endbibitem {}\fi
\ifx \bconflocation  \undefined \def \bconflocation#1{#1}\fi
\ifx \arxivurl  \undefined \def \arxivurl#1{\textsf{#1}}\fi
\csname PreBibitemsHook\endcsname

\bibitem{white_electronic_1981}
\begin{barticle}
\bauthor{\bsnm{White}, \binits{S.R.}},
\bauthor{\bsnm{Sham}, \binits{L.J.}}:
\batitle{Electronic properties of flat-band semiconductor heterostructures}.
\bjtitle{Phys. Rev. Lett.}
\bvolume{47}(\bissue{12}),
\bfpage{879}--\blpage{882}
(\byear{1981})
\end{barticle}
\endbibitem

\bibitem{sutherland_simple_1986}
\begin{barticle}
\bauthor{\bsnm{Sutherland}, \binits{B.}}:
\batitle{Simple system with quasiperiodic dynamics: {A} spin in a magnetic
  field}.
\bjtitle{Phys. Rev. Lett.}
\bvolume{57}(\bissue{6}),
\bfpage{770}--\blpage{773}
(\byear{1986})
\end{barticle}
\endbibitem

\bibitem{lieb_two_1989}
\begin{barticle}
\bauthor{\bsnm{Lieb}, \binits{E.H.}}:
\batitle{Two theorems on the {Hubbard} model}.
\bjtitle{Phys. Rev. Lett.}
\bvolume{62}(\bissue{10}),
\bfpage{1201}--\blpage{1204}
(\byear{1989})
\end{barticle}
\endbibitem

\bibitem{Kopnin2011}
\begin{barticle}
\bauthor{\bsnm{Kopnin}, \binits{N.B.}},
\bauthor{\bsnm{Heikkilä}, \binits{T.T.}},
\bauthor{\bsnm{Volovik}, \binits{G.E.}}:
\batitle{High-temperature surface superconductivity in topological flat-band
  systems}.
\bjtitle{Phys. Rev. B}
\bvolume{83}(\bissue{22}),
\bfpage{220503}
(\byear{2011})
\end{barticle}
\endbibitem

\bibitem{tang_strain-induced_2014}
\begin{barticle}
\bauthor{\bsnm{Tang}, \binits{E.}},
\bauthor{\bsnm{Fu}, \binits{L.}}:
\batitle{Strain-induced partially flat band, helical snake states and interface
  superconductivity in topological crystalline insulators}.
\bjtitle{Nat. Phys.}
\bvolume{10}(\bissue{12}),
\bfpage{964}--\blpage{969}
(\byear{2014})
\end{barticle}
\endbibitem

\bibitem{iglovikov_superconducting_2014}
\begin{barticle}
\bauthor{\bsnm{Iglovikov}, \binits{V.I.}},
\bauthor{\bsnm{Hébert}, \binits{F.}},
\bauthor{\bsnm{Grémaud}, \binits{B.}},
\bauthor{\bsnm{Batrouni}, \binits{G.G.}},
\bauthor{\bsnm{Scalettar}, \binits{R.T.}}:
\batitle{Superconducting transitions in flat-band systems}.
\bjtitle{Phys. Rev. B}
\bvolume{90}(\bissue{9}),
\bfpage{094506}
(\byear{2014})
\end{barticle}
\endbibitem

\bibitem{volovik_graphite_2018}
\begin{barticle}
\bauthor{\bsnm{Volovik}, \binits{G.E.}}:
\batitle{Graphite, graphene and the flat band superconductivity}.
\bjtitle{JETP Lett.}
\bvolume{107}(\bissue{8}),
\bfpage{516}--\blpage{517}
(\byear{2018})
\end{barticle}
\endbibitem

\bibitem{yin_negative_2019}
\begin{barticle}
\bauthor{\bsnm{Yin}, \binits{J.-X.}},
\bauthor{\bsnm{Zhang}, \binits{S.S.}},
\bauthor{\bsnm{Chang}, \binits{G.}},
\bauthor{\bsnm{Wang}, \binits{Q.}},
\bauthor{\bsnm{Tsirkin}, \binits{S.S.}},
\bauthor{\bsnm{Guguchia}, \binits{Z.}},
\bauthor{\bsnm{Lian}, \binits{B.}},
\bauthor{\bsnm{Zhou}, \binits{H.}},
\bauthor{\bsnm{Jiang}, \binits{K.}},
\bauthor{\bsnm{Belopolski}, \binits{I.}},
\bauthor{\bsnm{Shumiya}, \binits{N.}},
\bauthor{\bsnm{Multer}, \binits{D.}},
\bauthor{\bsnm{Litskevich}, \binits{M.}},
\bauthor{\bsnm{Cochran}, \binits{T.A.}},
\bauthor{\bsnm{Lin}, \binits{H.}},
\bauthor{\bsnm{Wang}, \binits{Z.}},
\bauthor{\bsnm{Neupert}, \binits{T.}},
\bauthor{\bsnm{Jia}, \binits{S.}},
\bauthor{\bsnm{Lei}, \binits{H.}},
\bauthor{\bsnm{Hasan}, \binits{M.Z.}}:
\batitle{Negative flat band magnetism in a spin–orbit-coupled correlated
  kagome magnet}.
\bjtitle{Nat. Phys.}
\bvolume{15}(\bissue{5}),
\bfpage{443}--\blpage{448}
(\byear{2019})
\end{barticle}
\endbibitem

\bibitem{liu_spectroscopy_2021}
\begin{barticle}
\bauthor{\bsnm{Liu}, \binits{X.}},
\bauthor{\bsnm{Chiu}, \binits{C.-L.}},
\bauthor{\bsnm{Lee}, \binits{J.Y.}},
\bauthor{\bsnm{Farahi}, \binits{G.}},
\bauthor{\bsnm{Watanabe}, \binits{K.}},
\bauthor{\bsnm{Taniguchi}, \binits{T.}},
\bauthor{\bsnm{Vishwanath}, \binits{A.}},
\bauthor{\bsnm{Yazdani}, \binits{A.}}:
\batitle{Spectroscopy of a tunable moir\'e system with a correlated and
  topological flat band}.
\bjtitle{Nat. Commun.}
\bvolume{12}(\bissue{1}),
\bfpage{2732}
(\byear{2021})
\end{barticle}
\endbibitem

\bibitem{rice_new_1975}
\begin{barticle}
\bauthor{\bsnm{Rice}, \binits{T.M.}},
\bauthor{\bsnm{Scott}, \binits{G.K.}}:
\batitle{New mechanism for a charge-density-wave instability}.
\bjtitle{Phys. Rev. Lett.}
\bvolume{35}(\bissue{2}),
\bfpage{120}--\blpage{123}
(\byear{1975})
\end{barticle}
\endbibitem

\bibitem{carpinelli_direct_1996}
\begin{barticle}
\bauthor{\bsnm{Carpinelli}, \binits{J.M.}},
\bauthor{\bsnm{Weitering}, \binits{H.H.}},
\bauthor{\bsnm{Plummer}, \binits{E.W.}},
\bauthor{\bsnm{Stumpf}, \binits{R.}}:
\batitle{Direct observation of a surface charge density wave}.
\bjtitle{Nature}
\bvolume{381}(\bissue{6581}),
\bfpage{398}--\blpage{400}
(\byear{1996})
\end{barticle}
\endbibitem

\bibitem{calandra_phonon-assisted_2018}
\begin{barticle}
\bauthor{\bsnm{Calandra}, \binits{M.}}:
\batitle{Phonon-assisted magnetic mott-insulating state in the charge density
  wave phase of single-layer 1{T}-{NbSe}$_2$}.
\bjtitle{Phys. Rev. Lett.}
\bvolume{121}(\bissue{2}),
\bfpage{026401}
(\byear{2018})
\end{barticle}
\endbibitem

\bibitem{wang_charge_2020}
\begin{barticle}
\bauthor{\bsnm{Wang}, \binits{W.}},
\bauthor{\bsnm{Wang}, \binits{B.}},
\bauthor{\bsnm{Gao}, \binits{Z.}},
\bauthor{\bsnm{Tang}, \binits{G.}},
\bauthor{\bsnm{Lei}, \binits{W.}},
\bauthor{\bsnm{Zheng}, \binits{X.}},
\bauthor{\bsnm{Li}, \binits{H.}},
\bauthor{\bsnm{Ming}, \binits{X.}},
\bauthor{\bsnm{Autieri}, \binits{C.}}:
\batitle{Charge density wave instability and pressure-induced superconductivity
  in bulk 1{T}-{Nb}{S}$_2$}.
\bjtitle{Phys. Rev. B}
\bvolume{102}(\bissue{15}),
\bfpage{155115}
(\byear{2020})
\end{barticle}
\endbibitem

\bibitem{miyahara_bcs_2007}
\begin{barticle}
\bauthor{\bsnm{Miyahara}, \binits{S.}},
\bauthor{\bsnm{Kusuta}, \binits{S.}},
\bauthor{\bsnm{Furukawa}, \binits{N.}}:
\batitle{{BCS} theory on a flat band lattice}.
\bjtitle{Physica C}
\bvolume{460-462},
\bfpage{1145}--\blpage{1146}
(\byear{2007})
\end{barticle}
\endbibitem

\bibitem{dora_lattice_2011}
\begin{barticle}
\bauthor{\bsnm{Dóra}, \binits{B.}},
\bauthor{\bsnm{Kailasvuori}, \binits{J.}},
\bauthor{\bsnm{Moessner}, \binits{R.}}:
\batitle{Lattice generalization of the {Dirac} equation to general spin and the
  role of the flat band}.
\bjtitle{Phys. Rev. B}
\bvolume{84}(\bissue{19}),
\bfpage{195422}
(\byear{2011})
\end{barticle}
\endbibitem

\bibitem{wang_nearly_2011}
\begin{barticle}
\bauthor{\bsnm{Wang}, \binits{F.}},
\bauthor{\bsnm{Ran}, \binits{Y.}}:
\batitle{Nearly flat band with {Chern} number {C}=2 on the dice lattice}.
\bjtitle{Phys. Rev. B}
\bvolume{84}(\bissue{24}),
\bfpage{241103}
(\byear{2011})
\end{barticle}
\endbibitem

\bibitem{bergholtz_topological_2013}
\begin{barticle}
\bauthor{\bsnm{Bergholtz}, \binits{E.J.}},
\bauthor{\bsnm{Liu}, \binits{Z.}}:
\batitle{Topological flat band models and fractional chern insulators}.
\bjtitle{Int. J. Mod. Phys. B}
\bvolume{27}(\bissue{24}),
\bfpage{1330017}
(\byear{2013})
\end{barticle}
\endbibitem

\bibitem{mukherjee_observation_2015}
\begin{barticle}
\bauthor{\bsnm{Mukherjee}, \binits{S.}},
\bauthor{\bsnm{Spracklen}, \binits{A.}},
\bauthor{\bsnm{Choudhury}, \binits{D.}},
\bauthor{\bsnm{Goldman}, \binits{N.}},
\bauthor{\bsnm{Öhberg}, \binits{P.}},
\bauthor{\bsnm{Andersson}, \binits{E.}},
\bauthor{\bsnm{Thomson}, \binits{R.R.}}:
\batitle{Observation of a localized flat-band state in a photonic {Lieb}
  lattice}.
\bjtitle{Phys. Rev. Lett.}
\bvolume{114}(\bissue{24}),
\bfpage{245504}
(\byear{2015})
\end{barticle}
\endbibitem

\bibitem{morales-inostroza_simple_2016}
\begin{barticle}
\bauthor{\bsnm{Morales-Inostroza}, \binits{L.}},
\bauthor{\bsnm{Vicencio}, \binits{R.A.}}:
\batitle{Simple method to construct flat-band lattices}.
\bjtitle{Phys. Rev. A}
\bvolume{94}(\bissue{4}),
\bfpage{043831}
(\byear{2016})
\end{barticle}
\endbibitem

\bibitem{julku_geometric_2016}
\begin{barticle}
\bauthor{\bsnm{Julku}, \binits{A.}},
\bauthor{\bsnm{Peotta}, \binits{S.}},
\bauthor{\bsnm{Vanhala}, \binits{T.I.}},
\bauthor{\bsnm{Kim}, \binits{D.-H.}},
\bauthor{\bsnm{Törmä}, \binits{P.}}:
\batitle{Geometric origin of superfluidity in the {Lieb}-lattice flat band}.
\bjtitle{Phys. Rev. Lett.}
\bvolume{117}(\bissue{4}),
\bfpage{045303}
(\byear{2016})
\end{barticle}
\endbibitem

\bibitem{mondaini_pairing_2018}
\begin{barticle}
\bauthor{\bsnm{Mondaini}, \binits{R.}},
\bauthor{\bsnm{Batrouni}, \binits{G.G.}},
\bauthor{\bsnm{Grémaud}, \binits{B.}}:
\batitle{Pairing and superconductivity in the flat band: {Creutz} lattice}.
\bjtitle{Phys. Rev. B}
\bvolume{98}(\bissue{15}),
\bfpage{155142}
(\byear{2018})
\end{barticle}
\endbibitem

\bibitem{maimaiti_compact_2017}
\begin{barticle}
\bauthor{\bsnm{Maimaiti}, \binits{W.}},
\bauthor{\bsnm{Andreanov}, \binits{A.}},
\bauthor{\bsnm{Park}, \binits{H.C.}},
\bauthor{\bsnm{Gendelman}, \binits{O.}},
\bauthor{\bsnm{Flach}, \binits{S.}}:
\batitle{Compact localized states and flat-band generators in one dimension}.
\bjtitle{Phys. Rev. B}
\bvolume{95}(\bissue{11}),
\bfpage{115135}
(\byear{2017})
\end{barticle}
\endbibitem

\bibitem{leykam_artificial_2018}
\begin{barticle}
\bauthor{\bsnm{Leykam}, \binits{D.}},
\bauthor{\bsnm{Andreanov}, \binits{A.}},
\bauthor{\bsnm{Flach}, \binits{S.}}:
\batitle{Artificial flat band systems: from lattice models to experiments}.
\bjtitle{Adv. Phys.:X}
\bvolume{3}(\bissue{1}),
\bfpage{1473052}
(\byear{2018})
\end{barticle}
\endbibitem

\bibitem{gardenier_p_2020}
\begin{barticle}
\bauthor{\bsnm{Gardenier}, \binits{T.S.}},
\bauthor{\bparticle{van~den} \bsnm{Broeke}, \binits{J.J.}},
\bauthor{\bsnm{Moes}, \binits{J.R.}},
\bauthor{\bsnm{Swart}, \binits{I.}},
\bauthor{\bsnm{Delerue}, \binits{C.}},
\bauthor{\bsnm{Slot}, \binits{M.R.}},
\bauthor{\bsnm{Smith}, \binits{C.M.}},
\bauthor{\bsnm{Vanmaekelbergh}, \binits{D.}}:
\batitle{$p$ orbital flat band and dirac cone in the electronic honeycomb
  lattice}.
\bjtitle{ACS Nano}
\bvolume{14}(\bissue{10}),
\bfpage{13638}--\blpage{13644}
(\byear{2020})
\end{barticle}
\endbibitem

\bibitem{chiu_fragile_2020}
\begin{barticle}
\bauthor{\bsnm{Chiu}, \binits{C.S.}},
\bauthor{\bsnm{Ma}, \binits{D.-S.}},
\bauthor{\bsnm{Song}, \binits{Z.-D.}},
\bauthor{\bsnm{Bernevig}, \binits{B.A.}},
\bauthor{\bsnm{Houck}, \binits{A.A.}}:
\batitle{Fragile topology in line-graph lattices with two, three, or four
  gapped flat bands}.
\bjtitle{Phys. Rev. Research}
\bvolume{2}(\bissue{4}),
\bfpage{043414}
(\byear{2020})
\end{barticle}
\endbibitem

\bibitem{ma_spin-orbit-induced_2020}
\begin{barticle}
\bauthor{\bsnm{Ma}, \binits{D.-S.}},
\bauthor{\bsnm{Xu}, \binits{Y.}},
\bauthor{\bsnm{Chiu}, \binits{C.S.}},
\bauthor{\bsnm{Regnault}, \binits{N.}},
\bauthor{\bsnm{Houck}, \binits{A.A.}},
\bauthor{\bsnm{Song}, \binits{Z.}},
\bauthor{\bsnm{Bernevig}, \binits{B.A.}}:
\batitle{Spin-orbit-induced topological flat bands in line and split graphs of
  bipartite lattices}.
\bjtitle{Phys. Rev. Lett.}
\bvolume{125}(\bissue{26}),
\bfpage{266403}
(\byear{2020})
\end{barticle}
\endbibitem

\bibitem{cadez_metal-insulator_2021}
\begin{barticle}
\bauthor{\bsnm{Čadež}, \binits{T.}},
\bauthor{\bsnm{Kim}, \binits{Y.}},
\bauthor{\bsnm{Andreanov}, \binits{A.}},
\bauthor{\bsnm{Flach}, \binits{S.}}:
\batitle{Metal-insulator transition in infinitesimally weakly disordered flat
  bands}.
\bjtitle{Phys. Rev. B}
\bvolume{104}(\bissue{18}),
\bfpage{180201}
(\byear{2021})
\end{barticle}
\endbibitem

\bibitem{provenzano_bringing_2020}
\begin{barticle}
\bauthor{\bsnm{Provenzano}, \binits{P.P.}}:
\batitle{Bringing order to the matrix}.
\bjtitle{Nat. Mater.}
\bvolume{19}(\bissue{2}),
\bfpage{130}--\blpage{131}
(\byear{2020})
\end{barticle}
\endbibitem

\bibitem{ghimire_competing_2020}
\begin{barticle}
\bauthor{\bsnm{Ghimire}, \binits{N.J.}},
\bauthor{\bsnm{Dally}, \binits{R.L.}},
\bauthor{\bsnm{Poudel}, \binits{L.}},
\bauthor{\bsnm{Jones}, \binits{D.C.}},
\bauthor{\bsnm{Michel}, \binits{D.}},
\bauthor{\bsnm{Magar}, \binits{N.T.}},
\bauthor{\bsnm{Bleuel}, \binits{M.}},
\bauthor{\bsnm{McGuire}, \binits{M.A.}},
\bauthor{\bsnm{Jiang}, \binits{J.S.}},
\bauthor{\bsnm{Mitchell}, \binits{J.F.}},
\bauthor{\bsnm{Lynn}, \binits{J.W.}},
\bauthor{\bsnm{Mazin}, \binits{I.I.}}:
\batitle{Competing magnetic phases and fluctuation-driven scalar spin chirality
  in the kagome metal {YMn}$_6${Sn}$_6$}.
\bjtitle{Sci. Adv.}
\bvolume{6}(\bissue{51}),
\bfpage{2680}
(\byear{2020})
\end{barticle}
\endbibitem

\bibitem{lou_charge-density-wave-induced_2022}
\begin{barticle}
\bauthor{\bsnm{Lou}, \binits{R.}},
\bauthor{\bsnm{Fedorov}, \binits{A.}},
\bauthor{\bsnm{Yin}, \binits{Q.}},
\bauthor{\bsnm{Kuibarov}, \binits{A.}},
\bauthor{\bsnm{Tu}, \binits{Z.}},
\bauthor{\bsnm{Gong}, \binits{C.}},
\bauthor{\bsnm{Schwier}, \binits{E.F.}},
\bauthor{\bsnm{Büchner}, \binits{B.}},
\bauthor{\bsnm{Lei}, \binits{H.}},
\bauthor{\bsnm{Borisenko}, \binits{S.}}:
\batitle{Charge-density-wave-induced peak-dip-hump structure and the multiband
  superconductivity in a kagome superconductor {CsV}$_3${Sb}$_5$}.
\bjtitle{Phys. Rev. Lett.}
\bvolume{128}(\bissue{3}),
\bfpage{036402}
(\byear{2022})
\end{barticle}
\endbibitem

\bibitem{neupert_charge_2022}
\begin{barticle}
\bauthor{\bsnm{Neupert}, \binits{T.}},
\bauthor{\bsnm{Denner}, \binits{M.M.}},
\bauthor{\bsnm{Yin}, \binits{J.-X.}},
\bauthor{\bsnm{Thomale}, \binits{R.}},
\bauthor{\bsnm{Hasan}, \binits{M.Z.}}:
\batitle{Charge order and superconductivity in kagome materials}.
\bjtitle{Nat. Phys.}
\bvolume{18}(\bissue{2}),
\bfpage{137}--\blpage{143}
(\byear{2022})
\end{barticle}
\endbibitem

\bibitem{mielke_ferromagnetism_1991}
\begin{barticle}
\bauthor{\bsnm{Mielke}, \binits{A.}}:
\batitle{Ferromagnetism in the {Hubbard} model on line graphs and further
  considerations}.
\bjtitle{J. Phys. A: Math. Gen.}
\bvolume{24}(\bissue{14}),
\bfpage{3311}--\blpage{3321}
(\byear{1991})
\end{barticle}
\endbibitem

\bibitem{mielke_ferromagnetic_1991}
\begin{barticle}
\bauthor{\bsnm{Mielke}, \binits{A.}}:
\batitle{Ferromagnetic ground states for the {Hubbard} model on line graphs}.
\bjtitle{J. Phys. A: Math. Gen.}
\bvolume{24}(\bissue{2}),
\bfpage{73}--\blpage{77}
(\byear{1991})
\end{barticle}
\endbibitem

\bibitem{chen_two-dimensional_2018}
\begin{barticle}
\bauthor{\bsnm{Chen}, \binits{Y.}},
\bauthor{\bsnm{Fan}, \binits{Z.}},
\bauthor{\bsnm{Zhang}, \binits{Z.}},
\bauthor{\bsnm{Niu}, \binits{W.}},
\bauthor{\bsnm{Li}, \binits{C.}},
\bauthor{\bsnm{Yang}, \binits{N.}},
\bauthor{\bsnm{Chen}, \binits{B.}},
\bauthor{\bsnm{Zhang}, \binits{H.}}:
\batitle{Two-dimensional metal nanomaterials: Synthesis, properties, and
  applications}.
\bjtitle{Chem. Rev.}
\bvolume{118}(\bissue{13}),
\bfpage{6409}--\blpage{6455}
(\byear{2018})
\end{barticle}
\endbibitem

\bibitem{jin_emerging_2018}
\begin{barticle}
\bauthor{\bsnm{Jin}, \binits{H.}},
\bauthor{\bsnm{Guo}, \binits{C.}},
\bauthor{\bsnm{Liu}, \binits{X.}},
\bauthor{\bsnm{Liu}, \binits{J.}},
\bauthor{\bsnm{Vasileff}, \binits{A.}},
\bauthor{\bsnm{Jiao}, \binits{Y.}},
\bauthor{\bsnm{Zheng}, \binits{Y.}},
\bauthor{\bsnm{Qiao}, \binits{S.-Z.}}:
\batitle{Emerging two-dimensional nanomaterials for electrocatalysis}.
\bjtitle{Chem. Rev.}
\bvolume{118}(\bissue{13}),
\bfpage{6337}--\blpage{6408}
(\byear{2018})
\end{barticle}
\endbibitem

\bibitem{han_van_2018}
\begin{barticle}
\bauthor{\bsnm{Han}, \binits{G.H.}},
\bauthor{\bsnm{Duong}, \binits{D.L.}},
\bauthor{\bsnm{Keum}, \binits{D.H.}},
\bauthor{\bsnm{Yun}, \binits{S.J.}},
\bauthor{\bsnm{Lee}, \binits{Y.H.}}:
\batitle{van der {Waals} metallic transition metal dichalcogenides}.
\bjtitle{Chem. Rev.}
\bvolume{118}(\bissue{13}),
\bfpage{6297}--\blpage{6336}
(\byear{2018})
\end{barticle}
\endbibitem

\bibitem{zeng_exploring_2018}
\begin{barticle}
\bauthor{\bsnm{Zeng}, \binits{M.}},
\bauthor{\bsnm{Xiao}, \binits{Y.}},
\bauthor{\bsnm{Liu}, \binits{J.}},
\bauthor{\bsnm{Yang}, \binits{K.}},
\bauthor{\bsnm{Fu}, \binits{L.}}:
\batitle{Exploring two-dimensional materials toward the next-generation
  circuits: From monomer design to assembly control}.
\bjtitle{Chem. Rev.}
\bvolume{118}(\bissue{13}),
\bfpage{6236}--\blpage{6296}
(\byear{2018})
\end{barticle}
\endbibitem

\bibitem{zhang_introduction_2018}
\begin{barticle}
\bauthor{\bsnm{Zhang}, \binits{H.}}:
\batitle{Introduction: {2D} materials chemistry}.
\bjtitle{Chem. Rev.}
\bvolume{118}(\bissue{13}),
\bfpage{6089}--\blpage{6090}
(\byear{2018})
\end{barticle}
\endbibitem

\bibitem{dong_interface-assisted_2018}
\begin{barticle}
\bauthor{\bsnm{Dong}, \binits{R.}},
\bauthor{\bsnm{Zhang}, \binits{T.}},
\bauthor{\bsnm{Feng}, \binits{X.}}:
\batitle{Interface-assisted synthesis of 2d materials: Trend and challenges}.
\bjtitle{Chem. Rev.}
\bvolume{118}(\bissue{13}),
\bfpage{6189}--\blpage{6235}
(\byear{2018})
\end{barticle}
\endbibitem

\bibitem{cai_chemical_2018}
\begin{barticle}
\bauthor{\bsnm{Cai}, \binits{Z.}},
\bauthor{\bsnm{Liu}, \binits{B.}},
\bauthor{\bsnm{Zou}, \binits{X.}},
\bauthor{\bsnm{Cheng}, \binits{H.-M.}}:
\batitle{Chemical vapor deposition growth and applications of two-dimensional
  materials and their heterostructures}.
\bjtitle{Chem. Rev.}
\bvolume{118}(\bissue{13}),
\bfpage{6091}--\blpage{6133}
(\byear{2018})
\end{barticle}
\endbibitem

\bibitem{li_epitaxial_2018}
\begin{barticle}
\bauthor{\bsnm{Li}, \binits{H.}},
\bauthor{\bsnm{Li}, \binits{Y.}},
\bauthor{\bsnm{Aljarb}, \binits{A.}},
\bauthor{\bsnm{Shi}, \binits{Y.}},
\bauthor{\bsnm{Li}, \binits{L.-J.}}:
\batitle{Epitaxial growth of two-dimensional layered transition-metal
  dichalcogenides: Growth mechanism, controllability, and scalability}.
\bjtitle{Chem. Rev.}
\bvolume{118}(\bissue{13}),
\bfpage{6134}--\blpage{6150}
(\byear{2018})
\end{barticle}
\endbibitem

\bibitem{jacqmin_direct_2014}
\begin{barticle}
\bauthor{\bsnm{Jacqmin}, \binits{T.}},
\bauthor{\bsnm{Carusotto}, \binits{I.}},
\bauthor{\bsnm{Sagnes}, \binits{I.}},
\bauthor{\bsnm{Abbarchi}, \binits{M.}},
\bauthor{\bsnm{Solnyshkov}, \binits{D.D.}},
\bauthor{\bsnm{Malpuech}, \binits{G.}},
\bauthor{\bsnm{Galopin}, \binits{E.}},
\bauthor{\bsnm{Lemaître}, \binits{A.}},
\bauthor{\bsnm{Bloch}, \binits{J.}},
\bauthor{\bsnm{Amo}, \binits{A.}}:
\batitle{Direct observation of dirac cones and a flat band in a honeycomb
  lattice for polaritons}.
\bjtitle{Phys. Rev. Lett.}
\bvolume{112}(\bissue{11}),
\bfpage{116402}
(\byear{2014})
\end{barticle}
\endbibitem

\bibitem{wang_flatbands_2016}
\begin{barticle}
\bauthor{\bsnm{Wang}, \binits{R.-N.}},
\bauthor{\bsnm{Zhang}, \binits{X.-R.}},
\bauthor{\bsnm{Wang}, \binits{S.-F.}},
\bauthor{\bsnm{Fu}, \binits{G.-S.}},
\bauthor{\bsnm{Wang}, \binits{J.-L.}}:
\batitle{Flatbands in {2D} boroxine-linked covalent organic frameworks}.
\bjtitle{Phys. Chem. Chem. Phys.}
\bvolume{18}(\bissue{2}),
\bfpage{1258}--\blpage{1264}
(\byear{2016})
\end{barticle}
\endbibitem

\bibitem{zhang_flat_2019}
\begin{barticle}
\bauthor{\bsnm{Zhang}, \binits{S.M.}},
\bauthor{\bsnm{Jin}, \binits{L.}}:
\batitle{Flat band in two-dimensional non-{Hermitian} optical lattices}.
\bjtitle{Phys. Rev. A}
\bvolume{100}(\bissue{4}),
\bfpage{043808}
(\byear{2019})
\end{barticle}
\endbibitem

\bibitem{crasto_de_lima_topological_2019}
\begin{barticle}
\bauthor{\bparticle{Crasto~de} \bsnm{Lima}, \binits{F.}},
\bauthor{\bsnm{Ferreira}, \binits{G.J.}},
\bauthor{\bsnm{Miwa}, \binits{R.H.}}:
\batitle{Topological flat band, {Dirac} fermions and quantum spin {Hall} phase
  in {2D} archimedean lattices}.
\bjtitle{Phys. Chem. Chem. Phys.}
\bvolume{21}(\bissue{40}),
\bfpage{22344}--\blpage{22350}
(\byear{2019})
\end{barticle}
\endbibitem

\bibitem{maimaiti_flat-band_2021}
\begin{barticle}
\bauthor{\bsnm{Maimaiti}, \binits{W.}},
\bauthor{\bsnm{Andreanov}, \binits{A.}},
\bauthor{\bsnm{Flach}, \binits{S.}}:
\batitle{Flat-band generator in two dimensions}.
\bjtitle{Phys. Rev. B}
\bvolume{103}(\bissue{16}),
\bfpage{165116}
(\byear{2021})
\end{barticle}
\endbibitem

\bibitem{geim_graphene_2009}
\begin{barticle}
\bauthor{\bsnm{Geim}, \binits{A.K.}}:
\batitle{Graphene: Status and prospects}.
\bjtitle{Science}
\bvolume{324}(\bissue{5934}),
\bfpage{1530}--\blpage{1534}
(\byear{2009})
\end{barticle}
\endbibitem

\bibitem{guo_graphene_2011}
\begin{botherref}
\oauthor{\bsnm{Guo}, \binits{B.}},
\oauthor{\bsnm{Fang}, \binits{L.}},
\oauthor{\bsnm{Zhang}, \binits{B.}},
\oauthor{\bsnm{Gong}, \binits{J.R.}}:
Graphene {Doping}: {A} {Review}.
Insciences J.,
80--89
(2011)
\end{botherref}
\endbibitem

\bibitem{lazar_adsorption_2013}
\begin{barticle}
\bauthor{\bsnm{Lazar}, \binits{P.}},
\bauthor{\bsnm{Karlický}, \binits{F.}},
\bauthor{\bsnm{Jurečka}, \binits{P.}},
\bauthor{\bsnm{Kocman}, \binits{M.}},
\bauthor{\bsnm{Otyepková}, \binits{E.}},
\bauthor{\bsnm{Šafářová}, \binits{K.}},
\bauthor{\bsnm{Otyepka}, \binits{M.}}:
\batitle{Adsorption of small organic molecules on graphene}.
\bjtitle{J. Am. Chem. Soc.}
\bvolume{135}(\bissue{16}),
\bfpage{6372}--\blpage{6377}
(\byear{2013})
\end{barticle}
\endbibitem

\bibitem{kong_molecular_2014}
\begin{barticle}
\bauthor{\bsnm{Kong}, \binits{L.}},
\bauthor{\bsnm{Enders}, \binits{A.}},
\bauthor{\bsnm{Rahman}, \binits{T.S.}},
\bauthor{\bsnm{Dowben}, \binits{P.A.}}:
\batitle{Molecular adsorption on graphene}.
\bjtitle{J. Phys.: Condens. Matter}
\bvolume{26}(\bissue{44}),
\bfpage{443001}
(\byear{2014})
\end{barticle}
\endbibitem

\bibitem{si_strain_2016}
\begin{barticle}
\bauthor{\bsnm{Si}, \binits{C.}},
\bauthor{\bsnm{Sun}, \binits{Z.}},
\bauthor{\bsnm{Liu}, \binits{F.}}:
\batitle{Strain engineering of graphene: {A} review}.
\bjtitle{Nanoscale}
\bvolume{8}(\bissue{6}),
\bfpage{3207}--\blpage{3217}
(\byear{2016})
\end{barticle}
\endbibitem

\bibitem{cao2018unconventional}
\begin{barticle}
\bauthor{\bsnm{Cao}, \binits{Y.}},
\bauthor{\bsnm{Fatemi}, \binits{V.}},
\bauthor{\bsnm{Fang}, \binits{S.}},
\bauthor{\bsnm{Watanabe}, \binits{K.}},
\bauthor{\bsnm{Taniguchi}, \binits{T.}},
\bauthor{\bsnm{Kaxiras}, \binits{E.}},
\bauthor{\bsnm{Jarillo-Herrero}, \binits{P.}}:
\batitle{Unconventional superconductivity in magic-angle graphene
  superlattices}.
\bjtitle{Nature}
\bvolume{556}(\bissue{7699}),
\bfpage{43}--\blpage{50}
(\byear{2018})
\end{barticle}
\endbibitem

\bibitem{PhysRevLett2018.121.087001}
\begin{barticle}
\bauthor{\bsnm{Xu}, \binits{C.}},
\bauthor{\bsnm{Balents}, \binits{L.}}:
\batitle{Topological superconductivity in twisted multilayer graphene}.
\bjtitle{Phys. Rev. Lett.}
\bvolume{121},
\bfpage{087001}
(\byear{2018})
\end{barticle}
\endbibitem

\bibitem{PhysRevB.98.085435}
\begin{barticle}
\bauthor{\bsnm{Zou}, \binits{L.}},
\bauthor{\bsnm{Po}, \binits{H.C.}},
\bauthor{\bsnm{Vishwanath}, \binits{A.}},
\bauthor{\bsnm{Senthil}, \binits{T.}}:
\batitle{Band structure of twisted bilayer graphene: Emergent symmetries,
  commensurate approximants, and wannier obstructions}.
\bjtitle{Phys. Rev. B}
\bvolume{98},
\bfpage{085435}
(\byear{2018})
\end{barticle}
\endbibitem

\bibitem{PhysRevB.98.085436}
\begin{barticle}
\bauthor{\bsnm{Fidrysiak}, \binits{M.}},
\bauthor{\bsnm{Zegrodnik}, \binits{M.}},
\bauthor{\bsnm{Spa\l{}ek}, \binits{J.}}:
\batitle{Unconventional topological superconductivity and phase diagram for an
  effective two-orbital model as applied to twisted bilayer graphene}.
\bjtitle{Phys. Rev. B}
\bvolume{98},
\bfpage{085436}
(\byear{2018})
\end{barticle}
\endbibitem

\bibitem{PhysRevX2018.8.031089}
\begin{barticle}
\bauthor{\bsnm{Po}, \binits{H.C.}},
\bauthor{\bsnm{Zou}, \binits{L.}},
\bauthor{\bsnm{Vishwanath}, \binits{A.}},
\bauthor{\bsnm{Senthil}, \binits{T.}}:
\batitle{Origin of {Mott} insulating behavior and superconductivity in twisted
  bilayer graphene}.
\bjtitle{Phys. Rev. X}
\bvolume{8},
\bfpage{031089}
(\byear{2018})
\end{barticle}
\endbibitem

\bibitem{PhysRevX2018.8.041041}
\begin{barticle}
\bauthor{\bsnm{Isobe}, \binits{H.}},
\bauthor{\bsnm{Yuan}, \binits{N.F.Q.}},
\bauthor{\bsnm{Fu}, \binits{L.}}:
\batitle{Unconventional superconductivity and density waves in twisted bilayer
  graphene}.
\bjtitle{Phys. Rev. X}
\bvolume{8},
\bfpage{041041}
(\byear{2018})
\end{barticle}
\endbibitem

\bibitem{PhysRevLett2018.121.257001}
\begin{barticle}
\bauthor{\bsnm{Wu}, \binits{F.}},
\bauthor{\bsnm{MacDonald}, \binits{A.H.}},
\bauthor{\bsnm{Martin}, \binits{I.}}:
\batitle{Theory of phonon-mediated superconductivity in twisted bilayer
  graphene}.
\bjtitle{Phys. Rev. Lett.}
\bvolume{121},
\bfpage{257001}
(\byear{2018})
\end{barticle}
\endbibitem

\bibitem{laksono2018singlet}
\begin{barticle}
\bauthor{\bsnm{Laksono}, \binits{E.}},
\bauthor{\bsnm{Leaw}, \binits{J.N.}},
\bauthor{\bsnm{Reaves}, \binits{A.}},
\bauthor{\bsnm{Singh}, \binits{M.}},
\bauthor{\bsnm{Wang}, \binits{X.}},
\bauthor{\bsnm{Adam}, \binits{S.}},
\bauthor{\bsnm{Gu}, \binits{X.}}:
\batitle{Singlet superconductivity enhanced by charge order in nested twisted
  bilayer graphene fermi surfaces}.
\bjtitle{Solid State Commun.}
\bvolume{282},
\bfpage{38}--\blpage{44}
(\byear{2018})
\end{barticle}
\endbibitem

\bibitem{PhysRevLett.121.217001}
\begin{barticle}
\bauthor{\bsnm{Liu}, \binits{C.-C.}},
\bauthor{\bsnm{Zhang}, \binits{L.-D.}},
\bauthor{\bsnm{Chen}, \binits{W.-Q.}},
\bauthor{\bsnm{Yang}, \binits{F.}}:
\batitle{Chiral spin density wave and $d+id$ superconductivity in the
  magic-angle-twisted bilayer graphene}.
\bjtitle{Phys. Rev. Lett.}
\bvolume{121},
\bfpage{217001}
(\byear{2018})
\end{barticle}
\endbibitem

\bibitem{PhysRevB.98.195101}
\begin{barticle}
\bauthor{\bsnm{Su}, \binits{Y.}},
\bauthor{\bsnm{Lin}, \binits{S.-Z.}}:
\batitle{Pairing symmetry and spontaneous vortex-antivortex lattice in
  superconducting twisted-bilayer graphene: Bogoliubov-de {Gennes} approach}.
\bjtitle{Phys. Rev. B}
\bvolume{98},
\bfpage{195101}
(\byear{2018})
\end{barticle}
\endbibitem

\bibitem{PhysRevB.98.220504}
\begin{barticle}
\bauthor{\bsnm{Peltonen}, \binits{T.J.}},
\bauthor{\bsnm{Ojaj\"arvi}, \binits{R.}},
\bauthor{\bsnm{Heikkil\"a}, \binits{T.T.}}:
\batitle{Mean-field theory for superconductivity in twisted bilayer graphene}.
\bjtitle{Phys. Rev. B}
\bvolume{98},
\bfpage{220504}
(\byear{2018})
\end{barticle}
\endbibitem

\bibitem{PhysRevB.98.241407}
\begin{barticle}
\bauthor{\bsnm{Kennes}, \binits{D.M.}},
\bauthor{\bsnm{Lischner}, \binits{J.}},
\bauthor{\bsnm{Karrasch}, \binits{C.}}:
\batitle{Strong correlations and $d+id$ superconductivity in twisted bilayer
  graphene}.
\bjtitle{Phys. Rev. B}
\bvolume{98},
\bfpage{241407}
(\byear{2018})
\end{barticle}
\endbibitem

\bibitem{guinea2018electrostatic}
\begin{barticle}
\bauthor{\bsnm{Guinea}, \binits{F.}},
\bauthor{\bsnm{Walet}, \binits{N.R.}}:
\batitle{Electrostatic effects, band distortions, and superconductivity in
  twisted graphene bilayers}.
\bjtitle{Proc. Natl. Acad. Sci. U.S.A.}
\bvolume{115}(\bissue{52}),
\bfpage{13174}--\blpage{13179}
(\byear{2018})
\end{barticle}
\endbibitem

\bibitem{PhysRevB.99.121407}
\begin{barticle}
\bauthor{\bsnm{Roy}, \binits{B.}},
\bauthor{\bparticle{Juri\ifmmode \check{c}\else
  \v{c}\fi{}i\ifmmode~\acute{c}\else} \bsnm{\'{c}\fi{}}, \binits{V.}}:
\batitle{Unconventional superconductivity in nearly flat bands in twisted
  bilayer graphene}.
\bjtitle{Phys. Rev. B}
\bvolume{99},
\bfpage{121407}
(\byear{2019})
\end{barticle}
\endbibitem

\bibitem{PhysRevLett.122.257002}
\begin{barticle}
\bauthor{\bsnm{Lian}, \binits{B.}},
\bauthor{\bsnm{Wang}, \binits{Z.}},
\bauthor{\bsnm{Bernevig}, \binits{B.A.}}:
\batitle{Twisted bilayer graphene: A phonon-driven superconductor}.
\bjtitle{Phys. Rev. Lett.}
\bvolume{122},
\bfpage{257002}
(\byear{2019})
\end{barticle}
\endbibitem

\bibitem{PhysRevLett.122.026801}
\begin{barticle}
\bauthor{\bsnm{Gonz\'alez}, \binits{J.}},
\bauthor{\bsnm{Stauber}, \binits{T.}}:
\batitle{Kohn-luttinger superconductivity in twisted bilayer graphene}.
\bjtitle{Phys. Rev. Lett.}
\bvolume{122},
\bfpage{026801}
(\byear{2019})
\end{barticle}
\endbibitem

\bibitem{PhysRevLett.122.246402}
\begin{barticle}
\bauthor{\bsnm{Seo}, \binits{K.}},
\bauthor{\bsnm{Kotov}, \binits{V.N.}},
\bauthor{\bsnm{Uchoa}, \binits{B.}}:
\batitle{Ferromagnetic {Mott} state in twisted graphene bilayers at the magic
  angle}.
\bjtitle{Phys. Rev. Lett.}
\bvolume{122},
\bfpage{246402}
(\byear{2019})
\end{barticle}
\endbibitem

\bibitem{yankowitz2019tuning}
\begin{barticle}
\bauthor{\bsnm{Yankowitz}, \binits{M.}},
\bauthor{\bsnm{Chen}, \binits{S.}},
\bauthor{\bsnm{Polshyn}, \binits{H.}},
\bauthor{\bsnm{Zhang}, \binits{Y.}},
\bauthor{\bsnm{Watanabe}, \binits{K.}},
\bauthor{\bsnm{Taniguchi}, \binits{T.}},
\bauthor{\bsnm{Graf}, \binits{D.}},
\bauthor{\bsnm{Young}, \binits{A.F.}},
\bauthor{\bsnm{Dean}, \binits{C.R.}}:
\batitle{Tuning superconductivity in twisted bilayer graphene}.
\bjtitle{Science}
\bvolume{363}(\bissue{6431}),
\bfpage{1059}--\blpage{1064}
(\byear{2019})
\end{barticle}
\endbibitem

\bibitem{lu2019superconductors}
\begin{barticle}
\bauthor{\bsnm{Lu}, \binits{X.}},
\bauthor{\bsnm{Stepanov}, \binits{P.}},
\bauthor{\bsnm{Yang}, \binits{W.}},
\bauthor{\bsnm{Xie}, \binits{M.}},
\bauthor{\bsnm{Aamir}, \binits{M.A.}},
\bauthor{\bsnm{Das}, \binits{I.}},
\bauthor{\bsnm{Urgell}, \binits{C.}},
\bauthor{\bsnm{Watanabe}, \binits{K.}},
\bauthor{\bsnm{Taniguchi}, \binits{T.}},
\bauthor{\bsnm{Zhang}, \binits{G.}}, \betal:
\batitle{Superconductors, orbital magnets and correlated states in magic-angle
  bilayer graphene}.
\bjtitle{Nature}
\bvolume{574}(\bissue{7780}),
\bfpage{653}--\blpage{657}
(\byear{2019})
\end{barticle}
\endbibitem

\bibitem{sharpe2019emergent}
\begin{barticle}
\bauthor{\bsnm{Sharpe}, \binits{A.L.}},
\bauthor{\bsnm{Fox}, \binits{E.J.}},
\bauthor{\bsnm{Barnard}, \binits{A.W.}},
\bauthor{\bsnm{Finney}, \binits{J.}},
\bauthor{\bsnm{Watanabe}, \binits{K.}},
\bauthor{\bsnm{Taniguchi}, \binits{T.}},
\bauthor{\bsnm{Kastner}, \binits{M.}},
\bauthor{\bsnm{Goldhaber-Gordon}, \binits{D.}}:
\batitle{Emergent ferromagnetism near three-quarters filling in twisted bilayer
  graphene}.
\bjtitle{Science}
\bvolume{365}(\bissue{6453}),
\bfpage{605}--\blpage{608}
(\byear{2019})
\end{barticle}
\endbibitem

\bibitem{huang2019antiferromagnetically}
\begin{barticle}
\bauthor{\bsnm{Huang}, \binits{T.}},
\bauthor{\bsnm{Zhang}, \binits{L.}},
\bauthor{\bsnm{Ma}, \binits{T.}}:
\batitle{Antiferromagnetically ordered mott insulator and $d + id$
  superconductivity in twisted bilayer graphene: A quantum monte carlo study}.
\bjtitle{Sci. Bull.}
\bvolume{64}(\bissue{5}),
\bfpage{310}--\blpage{314}
(\byear{2019})
\end{barticle}
\endbibitem

\bibitem{wu2018emergent}
\begin{botherref}
\oauthor{\bsnm{Wu}, \binits{X.-C.}},
\oauthor{\bsnm{Pawlak}, \binits{K.A.}},
\oauthor{\bsnm{Jian}, \binits{C.-M.}},
\oauthor{\bsnm{Xu}, \binits{C.}}:
Emergent superconductivity in the weak {Mott} insulator phase of bilayer
  graphene moir\`e superlattice.
arXiv:1805.06906
(2018)
\end{botherref}
\endbibitem

\bibitem{belsky_new_2002}
\begin{barticle}
\bauthor{\bsnm{Belsky}, \binits{A.}},
\bauthor{\bsnm{Hellenbrandt}, \binits{M.}},
\bauthor{\bsnm{Karen}, \binits{V.L.}},
\bauthor{\bsnm{Luksch}, \binits{P.}}:
\batitle{New developments in the inorganic crystal structure database ({ICSD}):
  {A}ccessibility in support of materials research and design}.
\bjtitle{Acta Crystallogr., Sect. B: Struct. Sci.}
\bvolume{58}(\bissue{3}),
\bfpage{364}--\blpage{369}
(\byear{2002})
\end{barticle}
\endbibitem

\bibitem{ashton_topology-scaling_2017}
\begin{barticle}
\bauthor{\bsnm{Ashton}, \binits{M.}},
\bauthor{\bsnm{Paul}, \binits{J.}},
\bauthor{\bsnm{Sinnott}, \binits{S.B.}},
\bauthor{\bsnm{Hennig}, \binits{R.G.}}:
\batitle{Topology-scaling identification of layered solids and stable
  exfoliated {2D} materials}.
\bjtitle{Phys. Rev. Lett.}
\bvolume{118}(\bissue{10}),
\bfpage{106101}
(\byear{2017})
\end{barticle}
\endbibitem

\bibitem{cheon_data_2017}
\begin{barticle}
\bauthor{\bsnm{Cheon}, \binits{G.}},
\bauthor{\bsnm{Duerloo}, \binits{K.-A.N.}},
\bauthor{\bsnm{Sendek}, \binits{A.D.}},
\bauthor{\bsnm{Porter}, \binits{C.}},
\bauthor{\bsnm{Chen}, \binits{Y.}},
\bauthor{\bsnm{Reed}, \binits{E.J.}}:
\batitle{Data mining for new two- and one-dimensional weakly bonded solids and
  lattice-commensurate heterostructures}.
\bjtitle{Nano Lett.}
\bvolume{17}(\bissue{3}),
\bfpage{1915}--\blpage{1923}
(\byear{2017})
\end{barticle}
\endbibitem

\bibitem{mounet_two-dimensional_2018}
\begin{barticle}
\bauthor{\bsnm{Mounet}, \binits{N.}},
\bauthor{\bsnm{Gibertini}, \binits{M.}},
\bauthor{\bsnm{Schwaller}, \binits{P.}},
\bauthor{\bsnm{Campi}, \binits{D.}},
\bauthor{\bsnm{Merkys}, \binits{A.}},
\bauthor{\bsnm{Marrazzo}, \binits{A.}},
\bauthor{\bsnm{Sohier}, \binits{T.}},
\bauthor{\bsnm{Castelli}, \binits{I.E.}},
\bauthor{\bsnm{Cepellotti}, \binits{A.}},
\bauthor{\bsnm{Pizzi}, \binits{G.}},
\bauthor{\bsnm{Marzari}, \binits{N.}}:
\batitle{Two-dimensional materials from high-throughput computational
  exfoliation of experimentally known compounds}.
\bjtitle{Nature}
\bvolume{13}(\bissue{3}),
\bfpage{246}--\blpage{252}
(\byear{2018})
\end{barticle}
\endbibitem

\bibitem{zhu_systematic_2018}
\begin{barticle}
\bauthor{\bsnm{Zhu}, \binits{Y.}},
\bauthor{\bsnm{Kong}, \binits{X.}},
\bauthor{\bsnm{Rhone}, \binits{T.D.}},
\bauthor{\bsnm{Guo}, \binits{H.}}:
\batitle{Systematic search for two-dimensional ferromagnetic materials}.
\bjtitle{Phys. Rev. Mater.}
\bvolume{2}(\bissue{8}),
\bfpage{081001}
(\byear{2018})
\end{barticle}
\endbibitem

\bibitem{calugaru_general_2022}
\begin{botherref}
\oauthor{\bsnm{Călugăru}, \binits{D.}},
\oauthor{\bsnm{Chew}, \binits{A.}},
\oauthor{\bsnm{Elcoro}, \binits{L.}},
\oauthor{\bsnm{Xu}, \binits{Y.}},
\oauthor{\bsnm{Regnault}, \binits{N.}},
\oauthor{\bsnm{Song}, \binits{Z.-D.}},
\oauthor{\bsnm{Bernevig}, \binits{B.A.}}:
General construction and topological classification of crystalline flat bands.
Nat. Phys.
\textbf{18}(2),
185--189
\end{botherref}
\endbibitem

\bibitem{bergman_band_2008}
\begin{botherref}
\oauthor{\bsnm{Bergman}, \binits{D.L.}},
\oauthor{\bsnm{Wu}, \binits{C.}},
\oauthor{\bsnm{Balents}, \binits{L.}}:
Band touching from real-space topology in frustrated hopping models.
Phys. Rev. B
\textbf{78}(12),
125104
\end{botherref}
\endbibitem

\bibitem{xu_theoretical_2020}
\begin{barticle}
\bauthor{\bsnm{Xu}, \binits{Q.-F.}},
\bauthor{\bsnm{Xie}, \binits{W.-Q.}},
\bauthor{\bsnm{Lu}, \binits{Z.-W.}},
\bauthor{\bsnm{Zhao}, \binits{Y.-J.}}:
\batitle{Theoretical study of enhanced ferromagnetism and tunable magnetic
  anisotropy of monolayer {CrI}$_3$ by surface adsorption}.
\bjtitle{Phys. Lett. A}
\bvolume{384}(\bissue{29}),
\bfpage{126754}
(\byear{2020})
\end{barticle}
\endbibitem

\bibitem{zhang_high-temperature_2020}
\begin{barticle}
\bauthor{\bsnm{Zhang}, \binits{H.}},
\bauthor{\bsnm{Yang}, \binits{W.}},
\bauthor{\bsnm{Ning}, \binits{Y.}},
\bauthor{\bsnm{Xu}, \binits{X.}}:
\batitle{High-temperature and multichannel quantum anomalous {Hall} effect in
  pristine and alkali–metal-doped {CrBr}$_3$ monolayers}.
\bjtitle{Nanoscale}
\bvolume{12}(\bissue{26}),
\bfpage{13964}--\blpage{13972}
(\byear{2020})
\end{barticle}
\endbibitem

\bibitem{regnault_catalogue_2022}
\begin{barticle}
\bauthor{\bsnm{Regnault}, \binits{N.}},
\bauthor{\bsnm{Xu}, \binits{Y.}},
\bauthor{\bsnm{Li}, \binits{M.-R.}},
\bauthor{\bsnm{Ma}, \binits{D.-S.}},
\bauthor{\bsnm{Jovanovic}, \binits{M.}},
\bauthor{\bsnm{Yazdani}, \binits{A.}},
\bauthor{\bsnm{Parkin}, \binits{S.S.P.}},
\bauthor{\bsnm{Felser}, \binits{C.}},
\bauthor{\bsnm{Schoop}, \binits{L.M.}},
\bauthor{\bsnm{Ong}, \binits{N.P.}},
\bauthor{\bsnm{Cava}, \binits{R.J.}},
\bauthor{\bsnm{Elcoro}, \binits{L.}},
\bauthor{\bsnm{Song}, \binits{Z.-D.}},
\bauthor{\bsnm{Bernevig}, \binits{B.A.}}:
\batitle{Catalogue of flat-band stoichiometric materials}.
\bjtitle{Nature}
\bvolume{603}(\bissue{7903}),
\bfpage{824}--\blpage{828}
(\byear{2022})
\end{barticle}
\endbibitem

\bibitem{PhysRevMaterials.5.084203}
\begin{barticle}
\bauthor{\bsnm{Liu}, \binits{H.}},
\bauthor{\bsnm{Meng}, \binits{S.}},
\bauthor{\bsnm{Liu}, \binits{F.}}:
\batitle{Screening two-dimensional materials with topological flat bands}.
\bjtitle{Phys. Rev. Mater.}
\bvolume{5},
\bfpage{084203}
(\byear{2021})
\end{barticle}
\endbibitem

\bibitem{miller_solid_1995}
\begin{botherref}
\oauthor{\bsnm{Miller}, \binits{G.J.}}:
Solid state chemistry of {Nb}$_3${Cl}$_8$: {Nb}$_3${TeCl}$_7$, mixed crystal
  formation, and intercalation.
J. Alloys Compd.,
8
(1995)
\end{botherref}
\endbibitem

\bibitem{huang_layer-dependent_2017}
\begin{barticle}
\bauthor{\bsnm{Huang}, \binits{B.}},
\bauthor{\bsnm{Clark}, \binits{G.}},
\bauthor{\bsnm{Navarro-Moratalla}, \binits{E.}},
\bauthor{\bsnm{Klein}, \binits{D.R.}},
\bauthor{\bsnm{Cheng}, \binits{R.}},
\bauthor{\bsnm{Seyler}, \binits{K.L.}},
\bauthor{\bsnm{Zhong}, \binits{D.}},
\bauthor{\bsnm{Schmidgall}, \binits{E.}},
\bauthor{\bsnm{McGuire}, \binits{M.A.}},
\bauthor{\bsnm{Cobden}, \binits{D.H.}},
\bauthor{\bsnm{Yao}, \binits{W.}},
\bauthor{\bsnm{Xiao}, \binits{D.}},
\bauthor{\bsnm{Jarillo-Herrero}, \binits{P.}},
\bauthor{\bsnm{Xu}, \binits{X.}}:
\batitle{Layer-dependent ferromagnetism in a van der {Waals} crystal down to
  the monolayer limit}.
\bjtitle{Nature}
\bvolume{546}(\bissue{7657}),
\bfpage{270}--\blpage{273}
(\byear{2017})
\end{barticle}
\endbibitem

\bibitem{lin_critical_2018}
\begin{barticle}
\bauthor{\bsnm{Lin}, \binits{G.T.}},
\bauthor{\bsnm{Luo}, \binits{X.}},
\bauthor{\bsnm{Chen}, \binits{F.C.}},
\bauthor{\bsnm{Yan}, \binits{J.}},
\bauthor{\bsnm{Gao}, \binits{J.J.}},
\bauthor{\bsnm{Sun}, \binits{Y.}},
\bauthor{\bsnm{Tong}, \binits{W.}},
\bauthor{\bsnm{Tong}, \binits{P.}},
\bauthor{\bsnm{Lu}, \binits{W.J.}},
\bauthor{\bsnm{Sheng}, \binits{Z.G.}},
\bauthor{\bsnm{Song}, \binits{W.H.}},
\bauthor{\bsnm{Zhu}, \binits{X.B.}},
\bauthor{\bsnm{Sun}, \binits{Y.P.}}:
\batitle{Critical behavior of two-dimensional intrinsically ferromagnetic
  semiconductor {CrI}$_3$}.
\bjtitle{Appl. Phys. Lett.}
\bvolume{112}(\bissue{7}),
\bfpage{072405}
(\byear{2018})
\end{barticle}
\endbibitem

\bibitem{shcherbakov_raman_2018}
\begin{barticle}
\bauthor{\bsnm{Shcherbakov}, \binits{D.}},
\bauthor{\bsnm{Stepanov}, \binits{P.}},
\bauthor{\bsnm{Weber}, \binits{D.}},
\bauthor{\bsnm{Wang}, \binits{Y.}},
\bauthor{\bsnm{Hu}, \binits{J.}},
\bauthor{\bsnm{Zhu}, \binits{Y.}},
\bauthor{\bsnm{Watanabe}, \binits{K.}},
\bauthor{\bsnm{Taniguchi}, \binits{T.}},
\bauthor{\bsnm{Mao}, \binits{Z.}},
\bauthor{\bsnm{Windl}, \binits{W.}},
\bauthor{\bsnm{Goldberger}, \binits{J.}},
\bauthor{\bsnm{Bockrath}, \binits{M.}},
\bauthor{\bsnm{Lau}, \binits{C.N.}}:
\batitle{Raman spectroscopy, photocatalytic degradation, and stabilization of
  atomically thin chromium tri-iodide}.
\bjtitle{Nano Lett.}
\bvolume{18}(\bissue{7}),
\bfpage{4214}--\blpage{4219}
(\byear{2018})
\end{barticle}
\endbibitem

\bibitem{thiel_probing_2019}
\begin{barticle}
\bauthor{\bsnm{Thiel}, \binits{L.}},
\bauthor{\bsnm{Wang}, \binits{Z.}},
\bauthor{\bsnm{Tschudin}, \binits{M.A.}},
\bauthor{\bsnm{Rohner}, \binits{D.}},
\bauthor{\bsnm{Gutiérrez-Lezama}, \binits{I.}},
\bauthor{\bsnm{Ubrig}, \binits{N.}},
\bauthor{\bsnm{Gibertini}, \binits{M.}},
\bauthor{\bsnm{Giannini}, \binits{E.}},
\bauthor{\bsnm{Morpurgo}, \binits{A.F.}},
\bauthor{\bsnm{Maletinsky}, \binits{P.}}:
\batitle{Probing magnetism in {2D} materials at the nanoscale with single-spin
  microscopy}.
\bjtitle{Science}
\bvolume{364}(\bissue{6444}),
\bfpage{973}--\blpage{976}
(\byear{2019})
\end{barticle}
\endbibitem

\bibitem{kresse_efficiency_1996}
\begin{barticle}
\bauthor{\bsnm{Kresse}, \binits{G.}},
\bauthor{\bsnm{Furthmüller}, \binits{J.}}:
\batitle{Efficiency of ab-initio total energy calculations for metals and
  semiconductors using a plane-wave basis set}.
\bjtitle{Comput. Mater. Sci.}
\bvolume{6}(\bissue{1}),
\bfpage{15}--\blpage{50}
(\byear{1996})
\end{barticle}
\endbibitem

\bibitem{monkhorst_special_1976}
\begin{barticle}
\bauthor{\bsnm{Monkhorst}, \binits{H.J.}},
\bauthor{\bsnm{Pack}, \binits{J.D.}}:
\batitle{Special points for {Brillouin}-zone integrations}.
\bjtitle{Phys. Rev. B}
\bvolume{13}(\bissue{12}),
\bfpage{5188}--\blpage{5192}
(\byear{1976})
\end{barticle}
\endbibitem

\bibitem{blochl_projector_1994}
\begin{barticle}
\bauthor{\bsnm{Blöchl}, \binits{P.E.}}:
\batitle{Projector augmented-wave method}.
\bjtitle{Phys. Rev. B}
\bvolume{50}(\bissue{24}),
\bfpage{17953}--\blpage{17979}
(\byear{1994})
\end{barticle}
\endbibitem

\bibitem{kresse_efficient_1996}
\begin{barticle}
\bauthor{\bsnm{Kresse}, \binits{G.}},
\bauthor{\bsnm{Furthmüller}, \binits{J.}}:
\batitle{Efficient iterative schemes for \textit{ab initio} total-energy
  calculations using a plane-wave basis set}.
\bjtitle{Phys. Rev. B}
\bvolume{54}(\bissue{16}),
\bfpage{11169}--\blpage{11186}
(\byear{1996})
\end{barticle}
\endbibitem

\bibitem{perdew_generalized_1996}
\begin{barticle}
\bauthor{\bsnm{Perdew}, \binits{J.P.}},
\bauthor{\bsnm{Burke}, \binits{K.}},
\bauthor{\bsnm{Ernzerhof}, \binits{M.}}:
\batitle{Generalized gradient approximation made simple}.
\bjtitle{Phys. Rev. Lett.}
\bvolume{77}(\bissue{18}),
\bfpage{3865}--\blpage{3868}
(\byear{1996})
\end{barticle}
\endbibitem

\bibitem{kresse_ultrasoft_1999}
\begin{barticle}
\bauthor{\bsnm{Kresse}, \binits{G.}},
\bauthor{\bsnm{Joubert}, \binits{D.}}:
\batitle{From ultrasoft pseudopotentials to the projector augmented-wave
  method}.
\bjtitle{Phys. Rev. B}
\bvolume{59}(\bissue{3}),
\bfpage{1758}--\blpage{1775}
(\byear{1999})
\end{barticle}
\endbibitem

\bibitem{calderon_aflow_2015}
\begin{barticle}
\bauthor{\bsnm{Calderon}, \binits{C.E.}},
\bauthor{\bsnm{Plata}, \binits{J.J.}},
\bauthor{\bsnm{Toher}, \binits{C.}},
\bauthor{\bsnm{Oses}, \binits{C.}},
\bauthor{\bsnm{Levy}, \binits{O.}},
\bauthor{\bsnm{Fornari}, \binits{M.}},
\bauthor{\bsnm{Natan}, \binits{A.}},
\bauthor{\bsnm{Mehl}, \binits{M.J.}},
\bauthor{\bsnm{Hart}, \binits{G.}},
\bauthor{\bsnm{Buongiorno~Nardelli}, \binits{M.}},
\bauthor{\bsnm{Curtarolo}, \binits{S.}}:
\batitle{The {AFLOW} standard for high-throughput materials science
  calculations}.
\bjtitle{Comput. Mater. Sci.}
\bvolume{108},
\bfpage{233}--\blpage{238}
(\byear{2015})
\end{barticle}
\endbibitem

\end{thebibliography}


\newpage
\bmhead{Methods}

We performed a high-throughput computation to reveal the electronica properties of the 861 vdW materials. We use density functional theory (DFT) as implemented in the Vienna ab initio Simulation Package (VASP)\cite{kresse_efficiency_1996} to calculate the band structures of the 861 2D materials. The Projector Augmented Wave (PAW) scheme with the Perdew-Burke-Ernzerhof (PBE) generalized gradient approximation (GGA) exchange-correlation function is employed \cite{monkhorst_special_1976,blochl_projector_1994,kresse_efficient_1996,perdew_generalized_1996,kresse_ultrasoft_1999}. We only focus on the DFT calculations performed without spin-orbit coupling. We adopt the GGA+\textit{U} method for the calculations, which considers the on-site Hubbard \textit{U} value of \textit{d} or \textit{f} electron in the mean-field approximation\cite{calderon_aflow_2015}.

\backmatter





\bmhead{Acknowledgments}

The work is supported by the National Key R\&D Program of China (Grant No. 2020YFA0308800), the National Natural Science Foundation of China (Grants Nos. 11734003, 12061131002),  the Strategic Priority Research Program of Chinese Academy of Sciences (Grant No. XDB30000000) and
the China Postdoctoral Science Foundation (Grant Nos.2020M680011, 2021T140057).

\bmhead{Competing interests}

The authors declare no competing interests.

\bmhead{Corresponding authors}

Correspondence and requests for materials should be addressed to Yugui Yao or Da-Shuai Ma or Run-Wu Zhang.

\end{document}